\documentclass[aps,pra,twocolumn,showpacs,superscriptaddress]{revtex4-1} 
\usepackage[colorlinks=true,linkcolor=blue,citecolor=blue,urlcolor=blue]{hyperref}
\usepackage{dcolumn}
\usepackage{bm}
\usepackage{subfigure}
\usepackage{amsthm}
\usepackage[english]{babel}
\usepackage{titlesec}
\usepackage{amsmath}
\usepackage{amsfonts}
\usepackage{amssymb}
\usepackage{mathrsfs}
\usepackage{lipsum}
\usepackage[pdftex]{color,graphicx}
\usepackage{xcolor}

\begin{document}

\title{Subdiffusion in a 1D Anderson insulator with random dephasing: Finite-size scaling, Griffiths effects, and possible implications for many-body localization}
\author{Scott R. Taylor}
\affiliation{Abdus Salam International Centre for Theoretical Physics, Strada Costiera 11, 34151 Trieste, Italy}
\author{Antonello Scardicchio}
\affiliation{Abdus Salam International Centre for Theoretical Physics, Strada Costiera 11, 34151 Trieste, Italy}
\affiliation{INFN, Sezione di Trieste, Via Valerio 2, 34126 Trieste, Italy}

\date{\today}

\begin{abstract}
We study transport in a one-dimensional boundary-driven Anderson insulator (the XX spin chain with onsite disorder) with randomly positioned onsite dephasing, observing a transition from diffusive to subdiffusive spin transport below a critical density of sites with dephasing.
This model is intended to mimic the passage of an excitation through (many-body) insulating regions or ergodic bubbles, therefore providing a toy model for the diffusion-subdiffusion transition observed in the disordered Heisenberg model \cite{Znidaric2016Diffusive}. We also present the exact solution of a semiclassical model of conductors and insulators introduced in Ref.~\onlinecite{AgarwalAnomalousDiffusion}, which exhibits both diffusive and subdiffusive phases, and qualitatively reproduces the results of the quantum system. The critical properties of both models, when passing from diffusion to subdiffusion, are interpreted in terms of ``Griffiths effects". We show that the finite-size scaling comes from the interplay of three characteristic lengths: one associated with disorder (the localization length), one with dephasing, and the third with the percolation problem defining large, rare, insulating regions. We conjecture that the latter, which grows logarithmically with system size, may potentially be responsible for the fact that heavy-tailed resistance distributions typical of Griffiths effects have not been observed in subdiffusive interacting systems.
\end{abstract}
\maketitle

\section{Introduction}
\label{sec:intro}

It has long been known that non-interacting quantum systems possess a transportless phase in the presence of sufficiently strong disorder, a phenomenon known as Anderson localization \cite{Anderson1958,abrahams1979scaling,Evers2008Review}.
More recent work on one-dimensional, \emph{interacting} quantum systems with strong disorder suggests the existence of a many-body localized (MBL) phase in which all quasiparticle transport is suppressed and entanglement spreads only logarithmically fast in time
\cite{Basko:2006hh,oganesyan2007localization,Pal2010,de2013ergodicity,nandkishore2015many,abanin2017recent,imbrie2017local}.
In the thermalizing phase, preceding the localization transition at weak disorder, several numerical studies have found evidence for anomalous subdiffusive transport of particles \cite{AgarwalAnomalousDiffusion,Reichman2014Absence,Znidaric2016Diffusive,Schulz2020Phenomenology} and energy \cite{Varma2017Energy,Schulz2018Energy,Mendoza2019Asymmetry}, and in general a violation of the Wiedemann-Franz law.

Despite widespread interest \cite{Luitz2016Extended,Khait2016SpinTransport,Luitz2016Anomalous,Gopalakrishnan2017Noise,Luitz2019Multifractality,Rispoli2019Quantum,Schulz2020Phenomenology,DeRoeck2020Subdiffusion,Luitz2017Ergodic,Gopalakrishnan2020Dynamics}, the debate on the microscopic origin of this subdiffusion is still not definitively settled.
The prevailing theory, first proposed in Ref.~\onlinecite{AgarwalAnomalousDiffusion}, is that subdiffusion is caused by ``Griffiths effects'', where rare regions of exceptionally strong disorder result in bottlenecks that slow down transport.
The phenomenological picture in the case of DC transport is that the system may be modelled by a chain of independent random resistors with resistances $r_i$ distributed as $P(r) \propto r^{-\nu}$ at large $r$.
For $1 < \nu \leq 2$ the average of $r$ diverges and the total resistance, given by the sum of individual resistances $R = \sum_{n=1}^L r_n$, no longer has a well defined average.
In this regime $R$ is dominated by the largest $r_n$ in the chain, and so the typical value of the total resistance scales as $R \propto L^{1 / (\nu - 1)}$, indicating a breakdown of Ohm's law, $R \propto L$ \cite{Hulin1990Strongly}.
However, while there is evidence of Griffiths effects in the structure of slow operators in the subdiffusive phase \cite{Pancotti2018Almost}, the essential ingredient of heavy-tailed resistance distributions has not been observed in numerical studies of large systems, casting doubt on this as the true origin of subdiffusion in the paradigmatic toy model for MBL, the Heisenberg spin chain \cite{Schulz2020Phenomenology}.
Griffiths effects are also a key feature in theories of the MBL transition and its critical properties, with thermalization proposed to result from a runaway growth of thermal inclusions \cite{Roeck2017Stability,Luitz2017How,Goihl2019Exploration,Crowley2020Avalanche,AltmanTheory2015,Potter2015Universal,Zhang2016Many,dumitrescu2017scaling,thiery2017microscopically,Thiery2018Many,goremykina2019analytically,Morningstar2019Renormalization,dumitrescu2019kosterlitz}.

In this paper we introduce a microscopic quantum system with a diffusion-subdiffusion transition consistent with the Griffiths effects picture: the disordered XX spin chain with random onsite dephasing.
This model is an Anderson insulator in the absence of the dephasing terms, equivalent to a system of non-interacting particles hopping on a disordered lattice.
We also present a solvable semiclassical model of conductors and insulators, possessing a subdiffusive phase driven by Griffiths effects, that captures the essential physics of the microscopic model.
This model is an example of the random-resistor systems introduced in Ref.~\onlinecite{AgarwalAnomalousDiffusion} and discussed above.

We show that the finite-size corrections to the asymptotic behavior of the microscopic model (be it diffusive or subdiffusive) are regulated by the interplay of three characteristic lengths: a dephasing length, a localization length, and the size of the largest insulating clusters. We discuss the interplay of these lengths by making use of a resistance beta function and we show how the interacting case, the Heisenberg model with disorder discussed in Ref.~\onlinecite{Znidaric2016Diffusive}, shows a similar phenomenology.
This makes our results relevant for the study of the MBL transition, and potentially offers a resolution of the discrepancy between some of the predictions of the model of subdiffusion presented in Ref.~\onlinecite{AgarwalAnomalousDiffusion} and the distributions observed in the more recent Ref.~\onlinecite{Schulz2020Phenomenology}.
We discuss how our microscopic model could loosely mimic a many-body localizable system with Griffiths effects, using the random dephasing as a controllable substitute for the dissipation caused by interactions, although naturally our non-interacting model cannot capture a MBL transition.

The paper is organized as follows: in Section~\ref{sec:dephasing} we present the microscopic model with numerical results, including a discussion of its relevance to MBL systems and an analysis of finite-size effects; in Section~\ref{sec:classical} we explore the semiclassical model both analytically and numerically; and we discuss our conclusions in Section~\ref{sec:Disc}.

\section{Random dephasing model}
\label{sec:dephasing}

The model we consider is the one-dimensional disordered XX chain, driven at the boundaries with random onsite dephasing.
This system has the Hamiltonian:
\begin{equation}
\label{eq:H}
\mathcal{H} = \sum_{n=1}^{L-1} \left( \sigma_n^x \sigma_{n+1}^x + \sigma_n^y \sigma_{n+1}^y \right) + \sum_{n=1}^L h_n \sigma_n^z,
\end{equation}
where $\sigma^{\mu}_n$ are Pauli matrices and $h_n \in [-W, W]$ are independent uniformly distributed random variables.
The Jordan-Wigner transformation maps this Hamiltonian exactly to non-interacting spinless fermions hopping on a disordered lattice \cite{JordanWigner}, and a spin current in the XX model corresponds to a particle current in the fermionic language.
The driving and dephasing are described by the Lindblad master equation:
\begin{equation}
\label{eq:Lindblad}
\frac{\mathrm{d} \rho}{\mathrm{d} t} = i[\rho, \mathcal{H}] + \sum_{k=0}^{L+1} \left( [L_k \rho, L^{+}_k] + [L_k, \rho L^{+}_k] \right).
\end{equation}
The spin current is driven by the jump operators:
\begin{equation}
\label{eq:driving}
L_0 = \sqrt{2 \Gamma} \sigma^{+}_1 \; \text{ and } \; L_{L+1} = \sqrt{2 \Gamma} \sigma^{-}_L,
\end{equation}
and the onsite dephasing by the jump operators:
\begin{equation}
\label{eq:dephasing}
L_n = \sqrt{\gamma_n / 2} \sigma^z_n \:\: \text{for} \:\: n = 1,2,\ldots,L,
\end{equation}
where for each site $\gamma_n = 0$ with probability $p$ and $\gamma_n = \gamma$ with probability $1-p$.

Similar setups have been used to study transport in both non-interacting \cite{Znidaric2010Exact,Znidaric2013Transport,Varma2017Fractality,Schulz2020Phenomenology} and interacting \cite{Znidaric2010Dephasing,Znidaric2011Transport,Znidaric2016Diffusive,Schulz2018Energy,Mendoza2019Asymmetry, Schulz2020Phenomenology} quantum systems.
After solving the Lindblad equation to find the non-equilibrium steady state (NESS) for a given realization of the disorder and dephasing, one can calculate the spin current $j_{\infty}$ and in turn the resistance $R \propto 1/j_{\infty}$.
The spin current from site $n$ to site $n+1$ is given by the expectation value of the operator $j_n = 2 ( \sigma^{x}_{n} \sigma^{y}_{n+1} - \sigma^{y}_{n} \sigma^{x}_{n+1})$, as defined by the continuity equation for the local magnetization, and is independent of $n$ in the NESS.
The nature of the transport can then be determined by the scaling of the typical resistance with the system size, $R \propto L^{\beta}$, where $\beta=1$ indicates diffusion and $\beta>1$ indicates subdiffusion (localization is signalled by $R \propto \exp(L/\xi)$, with $\xi$ the localization length, implying a divergence of $\beta$).
Similarly, as discussed earlier, the distribution of resistances can reveal the mechanism for subdiffusion, with the Griffiths effects picture necessarily implying the existence of heavy-tailed distributions.

The advantage of studying this non-interacting model is that the NESS current can be found exactly by manipulating matrices with dimensions equal to the system size $L$, rather than $4^L$ as would be the case with the full many-body state space.
This allows for the efficient numerical solution of large systems with $L \sim 1000$, without the need for approximations based on matrix-product operator methods \cite{Prosen2008Third,Znidaric2010Exact,Znidaric2013Transport}.
Details of the numerical method can be found in Appendix~\ref{app:method}.
These large system sizes are essential when studying transport and localization phenomena in disordered quantum systems due to strong finite-size effects \cite{Znidaric2016Diffusive,Suntajs2019Quantum,Abanin2019Distinguishing,Panda2020Can}, and are beyond what is achievable in interacting systems even using approximate methods.

In the limit of no dephasing, $p=1$, the system is Anderson localized (i.e.\ an insulator), and the resistance grows with system size as $R \propto e^{L / \xi}$ \cite{Anderson1958,Znidaric2013Transport,Schulz2020Phenomenology}.
In the opposite limit with dephasing on every site, $p=0$, the system is a diffusive conductor with $R \propto L$ \cite{Znidaric2010Exact,Znidaric2013Transport}.
For intermediate $p$, the system is made up of a series of these insulating and conducting regions, and as $p$ becomes large there will be an increasing number of long insulating segments.
This results in regions of the system with exponentially large resistances, and one might therefore expect subdiffusive transport as described by the Griffiths effects picture.
We explore this argument more thoroughly in Section~\ref{sec:classical}.

The interplay of conducting and insulating inclusions has been the focus of numerous works, including studies of how a single ergodic region can thermalize an otherwise localized system \cite{Luitz2017How,Goihl2019Exploration,Crowley2020Avalanche}, and renormalization group studies of the MBL transition \cite{AltmanTheory2015,Potter2015Universal,Zhang2016Many,dumitrescu2017scaling,thiery2017microscopically,Thiery2018Many,goremykina2019analytically,Morningstar2019Renormalization,dumitrescu2019kosterlitz}.
In another work, subdiffusion due to Griffiths effects was studied in a toy model where a collection of Anderson insulators were coupled by random matrices \cite{Schiro2020Toy}.
To the best of the authors' knowledge, we are presenting the first exact analysis of transport in a large quantum system with many conducting and insulating regions, and by employing an open setup we can directly access DC transport properties as studied in similar works on interacting systems \cite{Znidaric2016Diffusive,Schulz2018Energy,Mendoza2019Asymmetry,Schulz2020Phenomenology}.

In our numerical study we use the parameters $\Gamma = 1$ and $\gamma = 0.2$, and for a fixed disorder strength $W = 1,2,3,4$ we vary the dephasing fraction $p$ to probe the different regimes of transport.
For each parameter combination we sample many realizations of the disorder and dephasing (a minimum of 5,000 realizations for $L < 256$, 500 for $L \geq 256 $, and 200 for $L = 1024$), and we ensure that at least 95\% of the realizations converge to the correct NESS.
We define a \emph{beta function} $\partial \ln R/\partial\ln L$ and also perform numerical fits to the median resistance $R_{\star}(L)$ to determine the asymptotic scaling exponent $\beta$, including finite-size corrections.
We examine the finite-size flow of $\beta$ using this beta function, and we also compute it for the interacting XXZ model studied in Ref.\ \onlinecite{Znidaric2016Diffusive} (the admittedly noisy data are extracted from that paper, and are presented in Fig.\ \ref{fig:XXZScaling}).

We find that different finite-size corrections match the data more accurately in different parameter regimes.
To study the finite-size flow to the asymptotic functional form, $R_{\star} = a L^{\beta}$, we define $x = \ln L$ and $y = \ln R_{\star}$, and we use a fit of the form $y = a + \beta x + b / x$ for $W > 1$ and $R_{\star} = a (1 + b/L) L^{\beta}$ for $W = 1$ (in both cases these forms outperform a simple fit to $R_{\star} = a L^{\beta}$ with the smallest system sizes omitted).
These regimes are summarized in Table~\ref{tab:FiniteSize}.
Applying different fits can change the values of $\beta$ and the location of a potential transition to subdiffusion.

\begin{table}
    \centering
    \begin{tabular}{c|c}
        Regime & Fitting function \\
        \hline
        \rule{0pt}{4ex} \begin{tabular}{@{}c@{}}  Weak disorder, $\ell < \xi$ \\ ($W \leq 1$)\end{tabular} & $R_{\star} = a (1 + b/L)L^{\beta}$ \\
        \rule{0pt}{4ex} \begin{tabular}{@{}c@{}}  Strong disorder, $\xi < \ell$  \\ ($W > 1$)\end{tabular} & $\ln R_{\star} = a + \beta \ln L + b / \ln L$
    \end{tabular}
    \caption{Summary of the different resistance scaling fits used in the various regimes of the random dephasing model. The disorder strengths corresponding to each regime are indicated in parentheses.}
    \label{tab:FiniteSize}
\end{table}

\subsection{Relationship with the interacting model}
\label{sec:XXZ}

A key question is how the physics of this dephasing model is relevant to subdiffusion in interacting systems such as the disordered Heisenberg model.
In a many-body localizable system, Griffiths effects would be generated by complicated interactions between particles, and the presence or absence of rare insulating regions could only be inferred by measurements of related physical observables.
In the random dephasing model the insulating and thermal regions are introduced in a simple and controlled way by means of dephasing operators (see the explanation below), so this study may provide insight into the nature and origins of finite-size effects that one might observe in an interacting system with Griffiths effect.

\begin{figure}
    \centering
    \includegraphics[width=\columnwidth]{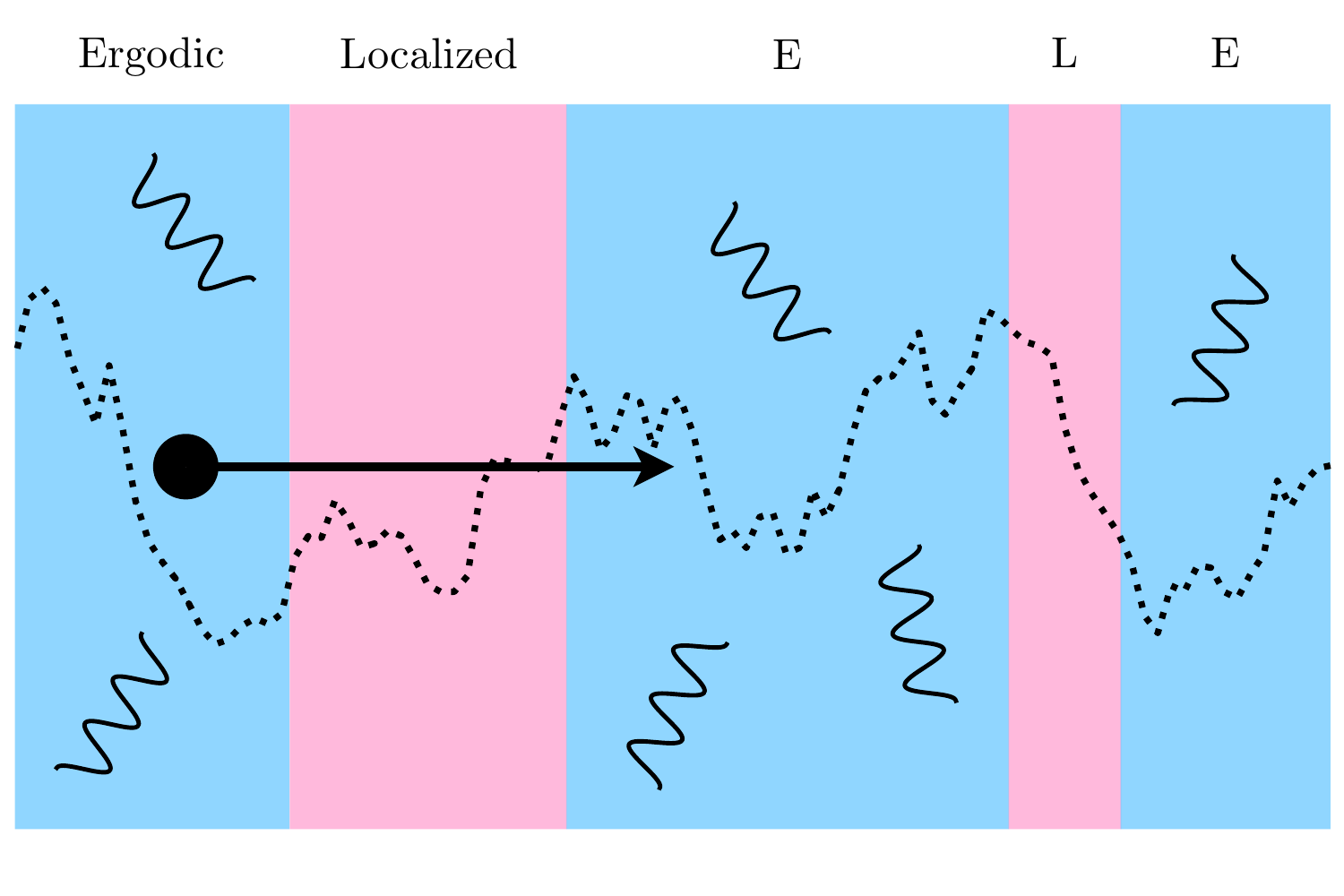}
    \caption{A toy model connecting our random dephasing model with the theory of a particle propagating on a random background, and in a sequence of ``ergodic bubbles" and localized regions (the blue and pink regions respectively). In the dephasing model, whether a site belongs to an ergodic or localized region is chosen randomly with probability $p$, which defines the distribution of sizes of the ergodic regions.}
    \label{fig:toy_model}
\end{figure}

In order to build an approximate mapping, consider that in the absence of dephasing the XX chain is simply the Heisenberg model in the limit of no interactions, so single-particle excitations move independently from one another. When the interactions are included, apart from a renormalization of the hopping and of the value of the disorder (which do not qualitatively change the motion in one dimension), from the point of view of a single excitation, a qualitatively new phenomenon can occur: If the particle goes through an ``ergodic bubble" (a cluster of sites that is locally thermal) \cite{Agarwal2017Rare,Thiery2018Many,dumitrescu2019kosterlitz} it dephases and can exchange energy with its surroundings, while if it passes through a localized region it does not dephase and effectively propagates without disturbance (see Fig.~\ref{fig:toy_model}).
By ``dephasing" here we mean that the particle acquires a random phase which depends on the state of the other particles in the system at the moment when the particle passes through the bubble. If we average over the random phase, we go from a unitary to a Lindblad equation of the kind studied in the present paper. As we will show in this paper, the dephasing needs to exceed a certain threshold, which depends on the size of the ergodic bubbles and the strength of the disorder, in order to turn a localized particle into a delocalized one.

Of course, in the full interacting model the problem has to be treated self-consistently: the rate of dephasing, the strength of the disorder, the size of the ergodic bubbles, and the effective hopping all depend on a few microscopic quantities defining the model (one in the Heisenberg model: $W/J$). The situation we have here, with random phases but the particles localized everywhere, is not self-consistent. Localization appears together with the disappearance of dephasing, as it is found in the distribution of the imaginary part of the self-energies \cite{Basko:2006hh}.

With this toy picture in mind, we see that on the thermal side of the MBL transition, the relatively small insulating regions of strong disorder in an otherwise thermal background are the cause of the subdiffusive transport (in accord with the Griffiths effects hypothesis).
We may, in this light, reexamine results from earlier studies on interacting disordered quantum systems for comparison with the random dephasing model.

Work on the scaling of resistance with system size in the disordered Heisenberg model has failed to find definitive evidence for Griffiths effects being the cause of subdiffusive transport (i.e.\ subdiffusive scaling of the resistance was observed but the resistance distributions did not have heavy tails) \cite{Schulz2020Phenomenology}.
This may be due to strong finite-size effects, as it is known that large systems are required to observe the asymptotic transport properties in interacting systems \cite{Znidaric2016Diffusive}.
In Ref.~\onlinecite{Schulz2020Phenomenology} it was shown that accurately simulating subdiffusive dynamics requires high bond dimensions in the time-evolving block decimation (TEBD) algorithm, so characterizing the subdiffusion in a large system has a restrictively high computational cost.
These TEBD studies on subdiffusion in large systems \cite{Znidaric2016Diffusive,Schulz2018Energy,Mendoza2019Asymmetry,Schulz2020Phenomenology} are limited to $L \lesssim 100$ in the subdiffusive phase, with the maximum achievable $L$ decreasing as the disorder strength increases and the transport becomes slower.
We will also find subdiffusive resistance scaling without heavy-tailed resistance distributions in some parameter regimes of the random dephasing model.

\begin{figure}
    \centering
    \includegraphics[width=\columnwidth]{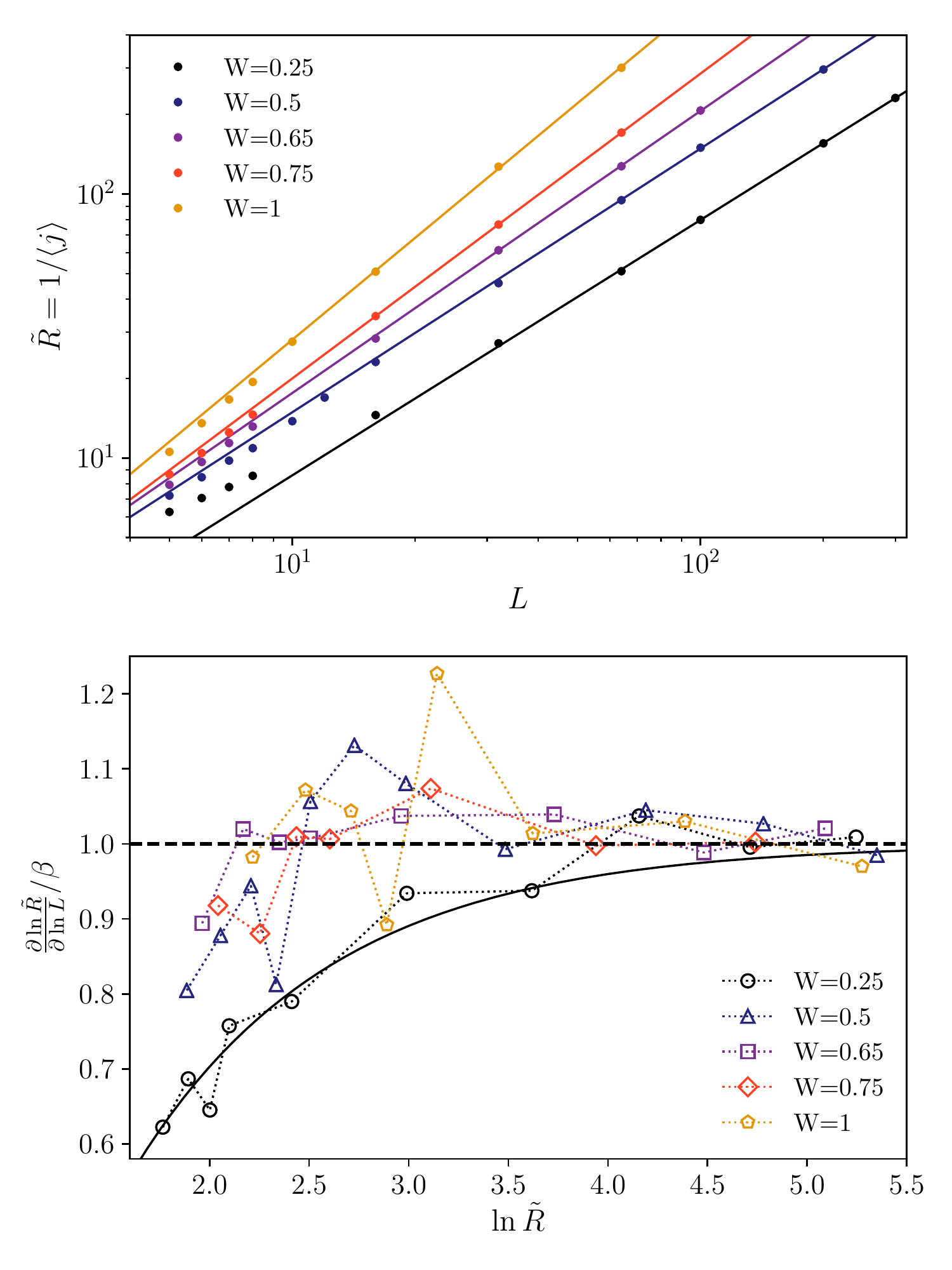}
    \caption{Scaling properties of the inverse NESS current with system size in the disordered Heisenberg spin chain.
    Upper panel: $\tilde{R}$ as a function of $L$ for several disorder strengths in the thermalizing phase. The points show the numerical data, and the lines indicate a fit of the form $ \tilde{R} = a L^{\beta}$ using the three largest system sizes. 
    Lower panel: The (discrete) beta function of the same data, normalized by its asymptotic value at $L \to \infty$. The solid black line indicates a fit of the form \eqref{eq:betasmallW} to the $W = 0.25$ data.
    Data reproduced from Ref.~[\onlinecite{Znidaric2016Diffusive}], for comparison with Fig.\ \ref{fig:betaf}.}
    \label{fig:XXZScaling}
\end{figure}

We will also observe similarities between the finite-size scaling of the resistance in the interacting system and in our dephasing model.
The scaling properties of the NESS current with system size in the thermalizing phase of the disordered Heisenberg spin chain were first presented in Ref.~\onlinecite{Znidaric2016Diffusive}, and we have reproduced the results in Fig.~\ref{fig:XXZScaling} (we maintain the convention from the original paper of stating disorder strength in relation to onsite fields that couple to spin operators, not Pauli matrices, so the diffusion-subdiffusion transition occurs at $W_c \approx 0.55$).
In the original paper the authors examined the average current $\langle j \rangle$ rather than the resistance, but the quantity $\tilde{R} = 1 / \langle j \rangle$ should behave in the same way as the typical (median) resistance, as the distribution of currents does not have a large tail.
In the upper panel of Fig.~\ref{fig:XXZScaling} the points show numerical data and the lines indicate fits of the form $\tilde{R} = a L^{\beta}$ evaluated on the largest three system sizes available.
For weak disorder $\tilde{R}$ approaches the asymptotic power-law scaling from above (note that the transport in the clean isotropic Heisenberg model is superdiffusive but not ballistic), while for stronger disorder it approaches the asymptotic scaling from below.
In the lower panel of Fig.~\ref{fig:XXZScaling} we show the resistance beta function, calculated using the discrete derivative, which we have normalized by its asymptotic value, $\beta$, to better compare the diffusive ($W \lesssim 0.55$) and subdiffusive ($W \gtrsim 0.55$) data.
For weak disorder the beta function approaches its asymptotic value from below (a fit of the form \eqref{eq:betasmallW} is shown for $W = 0.25$, see the discussion of finite-size effects with weak disorder in Section~\ref{sec:finitesize}), while for stronger disorder the asymptotic behavior is approached from above (these results are noisy because the TEBD algorithm is too computationally expensive to collect data as extensively as is possible for the non-interacting system).
In Section~\ref{sec:DephRes} we will see similar behavior for the random dephasing model, both in the scaling of the resistance with $L$ (in Fig.~\ref{fig:FiniteSize}) and the resistance beta function (in Fig.~\ref{fig:betaf}).
This suggests that the physics of the dephasing model, and therefore this work, may be relevant to fully interacting, disordered systems (which can eventually be many-body localized) with weak disorder.

\subsection{Finite-size effects: Three lengths}
\label{sec:finitesize}

\begin{figure}
    \centering
    \includegraphics[width=\columnwidth]{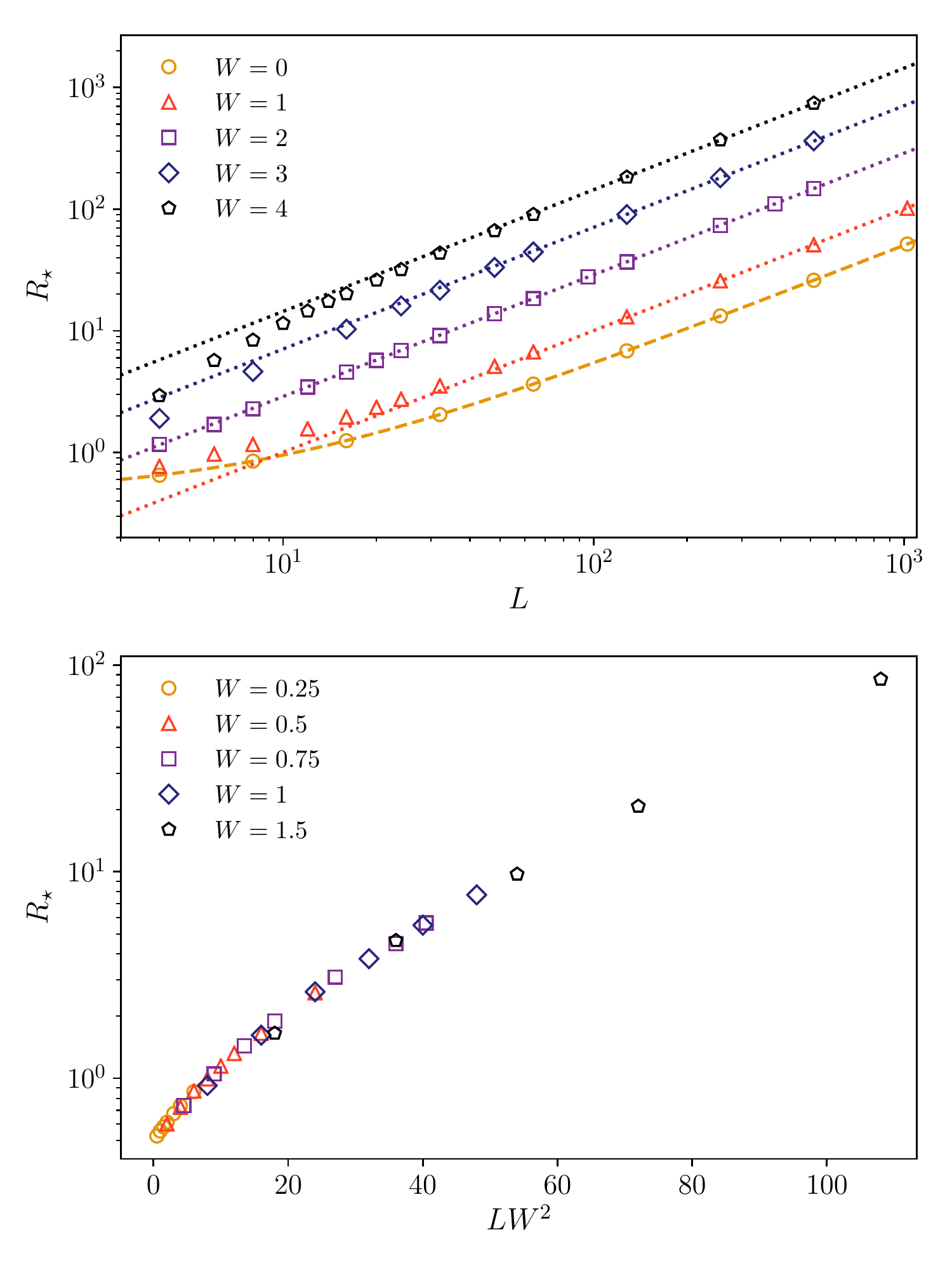}
    \caption{Finite-size effects in the resistance scaling. Upper panel: Median resistance $R_{\star}$ as a function of system size $L$ for several disorder strengths $W$ in a system with dephasing on every site ($p=0$). Dotted lines indicate the asymptotic diffusive scaling and the dashed line shows the exact result in the absence of disorder.
    Lower panel: Median resistance $R_{\star}$ as a function of the rescaled length $L W^2$ for several disorder strengths $W$ in a system with no dephasing ($p=1$), showing the scaling of the localization length at weak disorder $\xi \propto W^{-2}$.}
    \label{fig:FiniteSize}
\end{figure}

As described above, the behavior of the finite-size corrections are markedly different for $W=1$ and $W=4$, and different fitting functions work better for $R_\star(L)$. 
This is evident in Fig.~\ref{fig:FiniteSize}: the upper panel shows the scaling of the resistance with system size for a system with dephasing on every site, $p=0$, where for $W=1$ we see very similar behavior to that in the disorder-free case.
In the absence of dephasing and disorder the system exhibits ballistic transport, and the diffusion in the clean system with dephasing is a result of scattering due to the dephasing. At stronger disorder~(see Ref.~\onlinecite{Znidaric2013Transport}) we see a change of behavior, with $R$ approaching the asymptotic diffusive behavior from below, initially increasing faster than linearly with $L$.
Note the similarity with the results for the disordered Heisenberg model with weak and strong disorder, shown in the upper panel of Fig.~\ref{fig:XXZScaling}.
Our task is now to introduce length and resistance scales which separate the different behaviors.
These different flows with $L$ are caused by the interplay of three length scales: a length associated with the disorder strength $W$ (the localization length $\xi$), one with the density of dephasing sites $p$ (the largest cluster of sites without dephasing $s_\star$), and the third with the dephasing strength $\gamma$ (which we will call $\ell$).

Our first length is the localization length in 1d, which is known to be $\xi=\frac{24}{W^2}$ \cite{ThoulessReview,Izrailev1998Classical}. This is shown in the lower panel Figure \ref{fig:FiniteSize}, where for $LW^2\gtrsim 24$ the behavior of $R$ is indeed exponential, while for $LW^2\lesssim 24$ the behavior is power-law, $R\sim L^\beta$ with $\beta \simeq 0.12$ (in other words, to observe the exponential scaling of $R$ with system size we need $L > \xi$.). Notice that the law $\xi\sim 24/W^2$ is only valid for $W\lesssim 2$, while for large $W$ it is substituted by $\xi\sim 1/\ln W$ \cite{Pietracaprina2016Forward}.

Next, we will consider the largest ``insulator" size $s_\star$, which depends only on the density of dephasing sites $p$. From percolation theory in one dimension it is known that the typical value of the largest cluster of sites with \emph{no dephasing} is $s_\star=\log_{1/p}(L[1-p])$, to lowest order in $1/L$ \cite{Gordon1986Extreme}.
For the Griffiths picture to apply the resistance of these rare clusters must be exponentially large in $s_\star$ to create the power-law tail of $P(R)$.
The resistance of the insulating cluster must therefore be in its asymptotic scaling regime (i.e.\ the cluster is truly localized), and so we need $s_\star \gg \xi$.
If this condition is satisfied, then the contribution to the resistance of the largest cluster grows superlinearly with system size: $R_\star\propto e^{s_{\star}/\xi}=L^\beta$ with $\beta=1/(\xi\ln[1/p])$. 

The condition that the largest cluster is localized reads:
\begin{eqnarray}
    \log_{1/p}(L[1-p]) \gg \frac{24}{W^2}.
\end{eqnarray}
To see that this is not always satisfied in our numerics, consider the parameter combination $p=0.8, W=1, L=1024$: in this case $\xi=24$ and (the average) $s_\star=25$, so the largest cluster is about one localization length. The second largest cluster is on average $21$ lattice sites, so it is even smaller than a single localization length.
Moreover, the logarithmic dependence on $L$ means that if we want more than one localization length we must change $L$ enormously. Let us say we require 
\begin{eqnarray}
    \log_{1/p}(L[1-p]) > c \frac{24}{W^2}
\end{eqnarray}
with a minimum confidence of, say, $c=3$ (i.e.\ the largest cluster is at least 3 localization lengths). We see that
\begin{equation}
    L>\frac{1}{1-p}\left(\frac{1}{p}\right)^{72/W^2},
\end{equation}
and, for $p=0.8, W=1$ as before, this condition implies $L>10^7$. For $W=3$, on the contrary, for $p=0.5$ we get $L>500$, which is still within our reach. We therefore conclude that, even for the values of $L=10^3$ reached in our numerics, the data with $W=1$ are deep in the pre-asymptotic regime, while for $W=3,4$ the data are representative of the asymptotic behavior (for $p$ not too close to 1).
Clearly, an awareness of $s_{\star}$ is vital when trying to determine the asymptotic behavior of the system from numerical results.

In a given realization of the system, if the longest string of sites without dephasing, $s_0$, is larger than the localization length, $s_0\gtrsim \xi$, then we can show that the distribution of the resistance has a heavy power-law tail.
This can be seen by noting that the length of the longest insulating cluster $s_0$ obeys the Gumbel distribution for extreme values, with mode $s_{\star} = \log_{1/p} (L [1-p])$ and standard deviation $\pi /(\sqrt{6} \ln [1/p] )$ (this problem is equivalent to studying the longest run of consecutive heads when repeatedly tossing a biased coin) \cite{Gordon1986Extreme}.
Inserting these values into the Gumbel cumulative distribution function (CDF), we find the CDF for the length $s_0$ \cite{Gumbel2012Statistics}:
\begin{equation}
    \label{eq:GumbelCDF}
    F_{s_0}(s) = \exp \left( - \exp\left[(s_{\star}-s) \ln (1/p)\right]\right).
\end{equation}
Assuming that the resistance is dominated by this single long insulating cluster, and writing $q=e^{1/\xi},$ we have $R = R_1 q^{s_0}$, where $R_1$ is a constant.
The distribution function of the resistance is therefore:
\begin{equation}
    \label{eq:RDist}
    P(R) = \exp \left(-L[1-p] \left[ \frac{R_1}{R} \right]^{1/\beta} \right) \frac{L (1-p) R_1^{1/\beta}}{\beta R^{1+1/\beta}},
\end{equation}
where $\beta = \log_{1/p} (q)$ is the resistance scaling exponent.
For $R \gg R_1$, this distribution decays with a tail $P(R) \propto R^{-1-1/\beta}$, exactly as the Griffiths picture would predict for the subdiffusive scaling $R \propto L^{\beta}$.
If $\beta \leq 1$ then this argument does not hold, as the total resistance is not dominated by the longest insulating cluster.

However, if $s_0\lesssim \xi$, then we are in the small $L W^2$ region in the lower panel of Fig.~\ref{fig:FiniteSize} (say $s_0 W^2<24$). As discussed above, in this region the law $R=q^s$ is not valid: it is replaced by a law of the form $R\sim s^\beta$ where from the numerics $\beta \simeq 0.12$ (or at least $\beta \ll 1$). Using this relationship between $s$ and $R$, from \eqref{eq:GumbelCDF} we find that the distribution of $R$ decays like a stretched exponential
\begin{equation}
    \label{eq:StretchExp}
    P(R)\sim e^{-c R^{1/\beta}},
\end{equation}
faster than a power law. We will observe exactly this in the numerics discussed in Section~\ref{sec:DephRes}.

The third length, $\ell$, is associated with a string of consecutive sites with dephasing. We study this case in more detail and present numerics in Fig.~\ref{fig:FiniteSize}. We know that if dephasing is applied to every site of the chain, asymptotically one finds a resistance $R\propto L$. To a first approximation, if $\xi$ is large we can consider the situation in the absence of disorder. In this case the resistance of a chain of length $L$ with dephasing on every site has been calculated exactly in Ref.~\onlinecite{Znidaric2010Exact}:
\begin{equation}
    R = \frac{\gamma}{4} \left( 1 + \frac{\Gamma + \Gamma^{-1}-\gamma}{\gamma L} \right) L,
\end{equation}
which defines the asymptotic resistivity $\rho=\gamma/4$, or analogously the diffusion coefficient $D \propto 1/\gamma$.
From the relation $D = v \ell$, where $v$ is the velocity of excitations of the clean system (independent of $\gamma$), we see that $\ell \propto 1 / \gamma$. The same length dominates the finite-size effects for $\gamma\ll \Gamma$ ($\Gamma=1$ in our numerics) since we can write $R = \rho L \left( 1 + 2\frac{\ell}{L} \right)$.

For system sizes smaller than $\ell$, or resistances smaller than $R_0\equiv \rho \ell=\gamma\ell/4$, the resistance grows \emph{slower} than $L^1$, since the system goes from ballistic to diffusive transport. This can be seen by looking at the resistance beta function:
\begin{equation}
    \label{eq:betasmallW}
    \frac{\partial \ln R}{\partial \ln L}=1-\frac{R_0}{R}.
\end{equation}

On the other hand, if the disorder is much larger than the dephasing, $\xi\ll\ell$, for systems with size $L$ such that $\xi< L\ll \ell$ the resistance will scale exponentially with $L$. So, writing $R/R_1=e^{L/\xi}$, in this regime:
\begin{equation}
    \frac{\partial \ln R}{\partial \ln L}=\ln(R/R_1) > 1.
\end{equation}
For $L>\ell$, however, it must reach the condition $\frac{\partial \ln R}{\partial \ln L}\to 1$.

\begin{figure}
    \centering
    \includegraphics[width=\columnwidth]{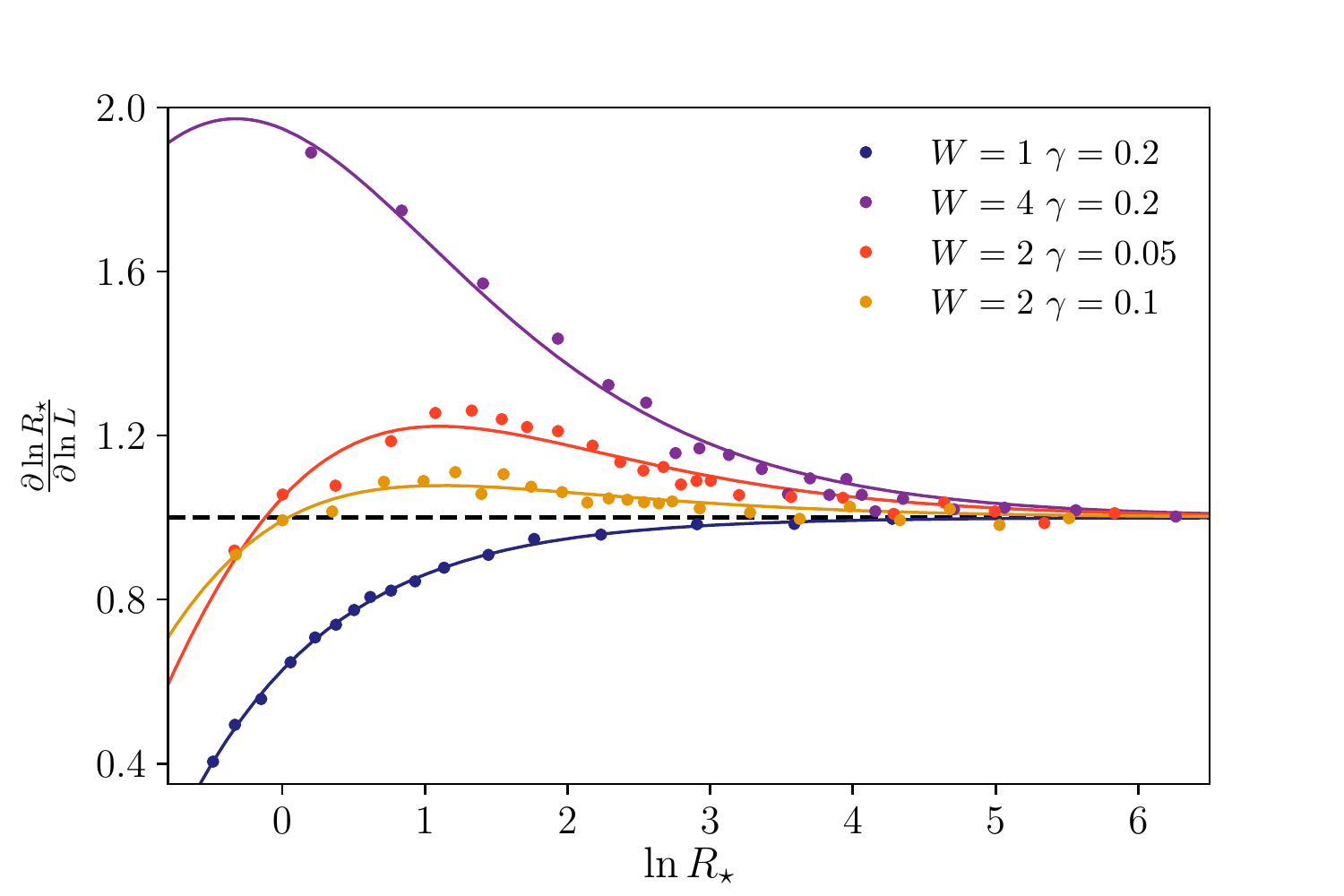}
    \caption{The resistance beta function for various disorder and dephasing values with $p=0$ (discrete derivatives of the data in Fig.~\ref{fig:FiniteSize}). We go from $W=4,\gamma=0.2$ which is large disorder and small dephasing to $W=1,\gamma=0.2$ which is small disorder. Intermediate cases $W=2,\gamma=0.05,0.1$ are shown. Numerical results are fitted with a phenomenological function of the form \eqref{eq:pheno_beta}, indicated by the lines of the corresponding color.}
    \label{fig:betaf}
\end{figure}

Putting everything together we see that we can distinguish two regimes, depending on whether we have $\xi\lesssim L\ll \ell$ or $\ell\ll L\lesssim \xi$ (or in terms of resistances, whether we have $R_0\ll R_1$ or $R_1\ll R_0$). Fixing $\gamma,$ we have large-disorder and small-disorder finite-size scaling behaviors which are completely different, as described in Table~\ref{tab:FiniteSize}. We find, however, that an extremely good, phenomenological, two-parameter fit function is given by
\begin{equation}
    \frac{\partial \ln R}{\partial \ln L}=1+\frac{\ln(R/R_a)}{1+R/R_b},
    \label{eq:pheno_beta}
\end{equation}
where $R_{a,b}$ are two fitting parameters. This form fits all the data we have for any $L,W,\gamma$ with good accuracy. The weak disorder case is obtained by $\ln(R_a/R^*) R_b=-R_0$ (for some $R^*$ of the size of the observed resistances) while the large disorder case comes from the region $R \sim R_a = R_1 \ll R_b$.
Fig.~\ref{fig:betaf} shows examples of the beta function from our numerical data (calculated using a discrete derivative), showing good agreement with the phenomenological form \eqref{eq:pheno_beta} in the strong disorder, strong dephasing, and intermediate regimes.
There are similarities between the results of Fig.~\ref{fig:betaf} and the beta function of the disordered Heisenberg model in Fig.~\ref{fig:XXZScaling}, with both approaching their asymptotic values from below in the case of weak disorder, and from above in the case of strong disorder.

We notice that the definition of this beta function is the same (except for an overall sign and the identification $g\propto 1/R$) as the typical conductance beta function which is amply described in the literature on disordered systems \cite{lee1985disordered}. It can be computed in perturbation theory in the weak localization regime and in the strongly localized regime for a variety of cases.
However, in the literature we have not found a discussion of this function in the setup of open system dynamics as presented above.

We are now ready to discuss the general scenario, with both random dephasing and disorder.

\subsection{Results: Diffusion-subdiffusion transition and critical point}
\label{sec:DephRes}

\begin{figure}
    \centering
    \includegraphics[width=\columnwidth]{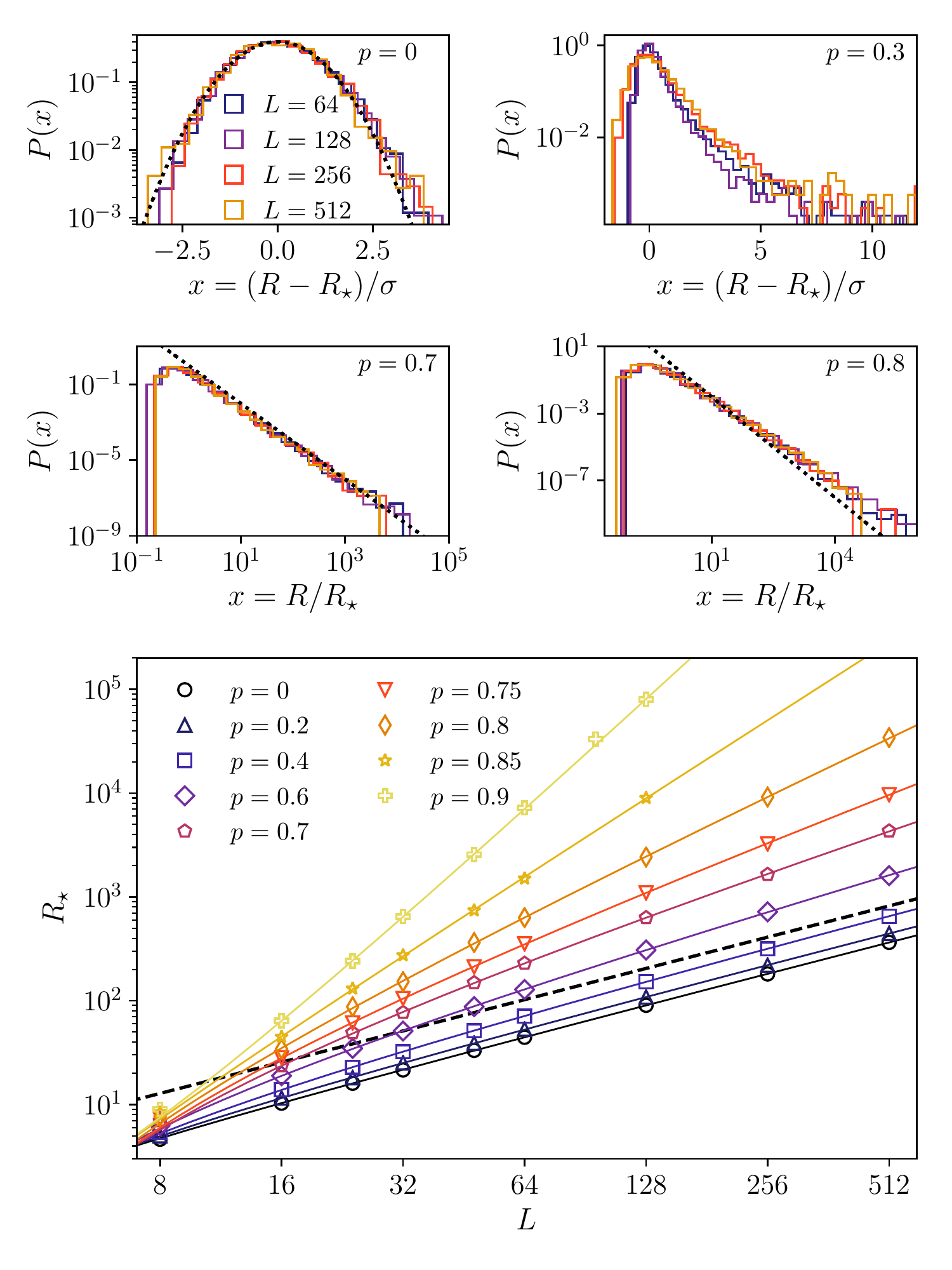}
    \caption{Numerical results on the resistance in the random dephasing model with $W=3$. Upper panels: Histograms of the rescaled resistance in the diffusive phase $p = 0$ (top left) and $p=0.3$ (top right), and the subdiffusive phase $p=0.7$ (middle left) and $p=0.8$ (middle right). The diffusive rescaling $(R-R_{\star})/\sigma$ is compared to a Gaussian distribution (black dotted line) for $p=0$. In the subdiffusive phase the resistance is rescaled as $R / R_{\star}$, and the black dotted lines indicate a $R^{-2}$ tail. Lower panel: Scaling of the median resistance $R_{\star}$ with system size $L$ for several dephasing fractions $p$. The lines in the corresponding color indicate the numerical fits to the data described in the text. The black dashed line indicates diffusive scaling $R_{\star} \propto L$.}
    \label{fig:hists}
\end{figure}

Fig.~\ref{fig:hists} summarizes the behavior of the resistance for a system with $W=3$.
The lower panel shows an example of the power-law scaling of the median resistance $R_{\star}$ with system size for several dephasing fractions $p$; the statistical uncertainties are smaller than the symbols and have therefore been omitted.
It is clear that for $p \lesssim 0.4$ the lines are parallel, indicating the same scaling with $L$ (we will show later that this corresponds to the diffusive phase), whereas for larger $p$ the resistance grows more steeply with an exponent that increases with $p$.
The black dashed line indicates the diffusive behavior $R_{\star} \propto L$.
Numerical fits to these data, including finite-size corrections as described above, are indicated by the lines.

Histograms of the resistance for various parameter combinations are shown in the upper panels of Fig.~\ref{fig:hists}.
We find that, in the diffusive phase, the resistance distributions $P(R)$ for different system sizes can be collapsed by a rescaling $(R - R_{\star}) / \sigma$, where $R_{\star} \propto L$ is the average or typical value and $\sigma \propto \sqrt{L}$ is the standard deviation or width.
Contrastingly, in the subdiffusive phase the collapse can be achieved by a rescaling of the form $R / R_{\star}$, indicating that both the typical value and the width of the distribution grow like $L^{\beta}$.
Deep in the diffusive phase the distribution is well approximated by a Gaussian (see the black dotted line on the $p = 0$ histogram), but as the system approaches the transition to subdiffusion a tail develops at large $R$ (see the $p = 0.3$  histogram).
In the subdiffusive phase we see heavy power-law tails in $P(R)$, as shown in the $p = 0.7$ and $p=0.8$ histograms (the black dotted lines indicate the $P(R) \propto R^{-2}$ tail that signals the onset of subdiffusion).

\begin{figure}
    \centering
    \includegraphics[width=\columnwidth]{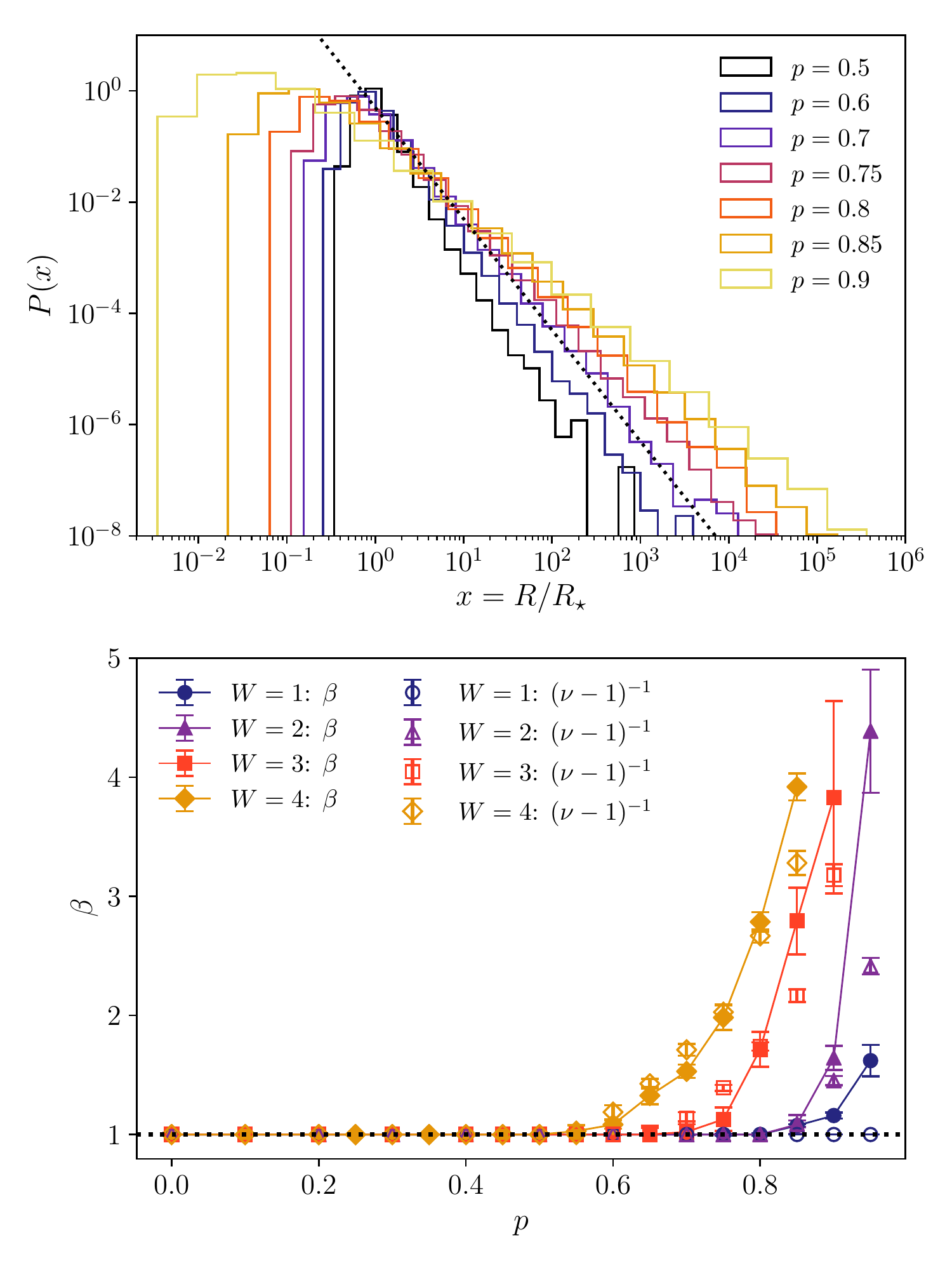}
    \caption{Upper panel: Histograms of the rescaled resistance $R / R_{\star}$ for $W = 3$, $L = 128$, and several dephasing fractions $p$. The black dotted line indicates a heavy $R^{-2}$ tail.
    Lower panel: Comparison of the resistance scaling exponent $\beta$ (connected filled points) with the predicted value based on the power-law tail of the histogram $(\nu - 1)^{-1}$ (unconnected unfilled points in the corresponding color). The discrepancy at $p\gtrsim 0.8$ and disorder $W\lesssim 3$ is attributed to the fact that the largest insulating cluster has a size $s_\star$ which is smaller than the localization length $\xi$, therefore the distribution of resistance does not show the appropriate power-law tail.}
    \label{fig:PLtails}
\end{figure}

The two phases can be described by the asymptotic behavior of the typical resistance $R_\star\propto L^\beta$:
\begin{equation}
    \begin{cases}
    \beta=1\ &\text{if } p<p_c,\\
    \beta>1\ &\text{if } p>p_c.
    \end{cases}
\end{equation}
In the lower panel of Fig.~\ref{fig:PLtails} the connected filled points show the scaling exponent $\beta$, found from a fit to the median resistance beta function (see discussion below) for $W = 2, 3, 4$, and from a direct fit of $R_{\star}$ as a function of $L$ for $W=1$ as described in Table~\ref{tab:FiniteSize}.
For each $W$ the transport is diffusive for small $p$ (i.e.\ $\beta = 1$), but upon increasing $p$ above a critical value $p_c(W)$ the transport becomes subdiffusive (i.e.\ $\beta > 1$).
The critical dephasing fraction $p_c(W)$ decreases with $W$, as the closed system is more strongly localized, and therefore the transport is weaker for a given $p$.

The upper panel of Fig.~\ref{fig:PLtails} shows histograms of $R / R_{\star}$ for a range of $p$ values with $L=128$ and $W=3$, and the black dashed line shows the $R^{-2}$ tail that signifies the onset of subdiffusion in the Griffiths effects picture.
We see that in the subdiffusive phase ($p \gtrsim 0.5$) the distribution tails decay more slowly than $R^{-2}$ and become heavier as the transport becomes slower.

If the Griffiths effects picture is correct, the exponent of the histogram decay $\nu$ should be related to the exponent of the resistance scaling as $\beta = (\nu - 1)^{-1}$.
The values of $(\nu - 1)^{-1}$ from the $L=128$ histograms are shown by the unfilled points in the lower panel.
We see that in the subdiffusive phase there is reasonable agreement for $W = 3$ and $4$ when $p$ is not too large, as discussed in Section~\ref{sec:finitesize}, while for weaker disorder and large $p$ the agreement becomes poor.

\begin{figure}
    \centering
    \includegraphics[width=\columnwidth]{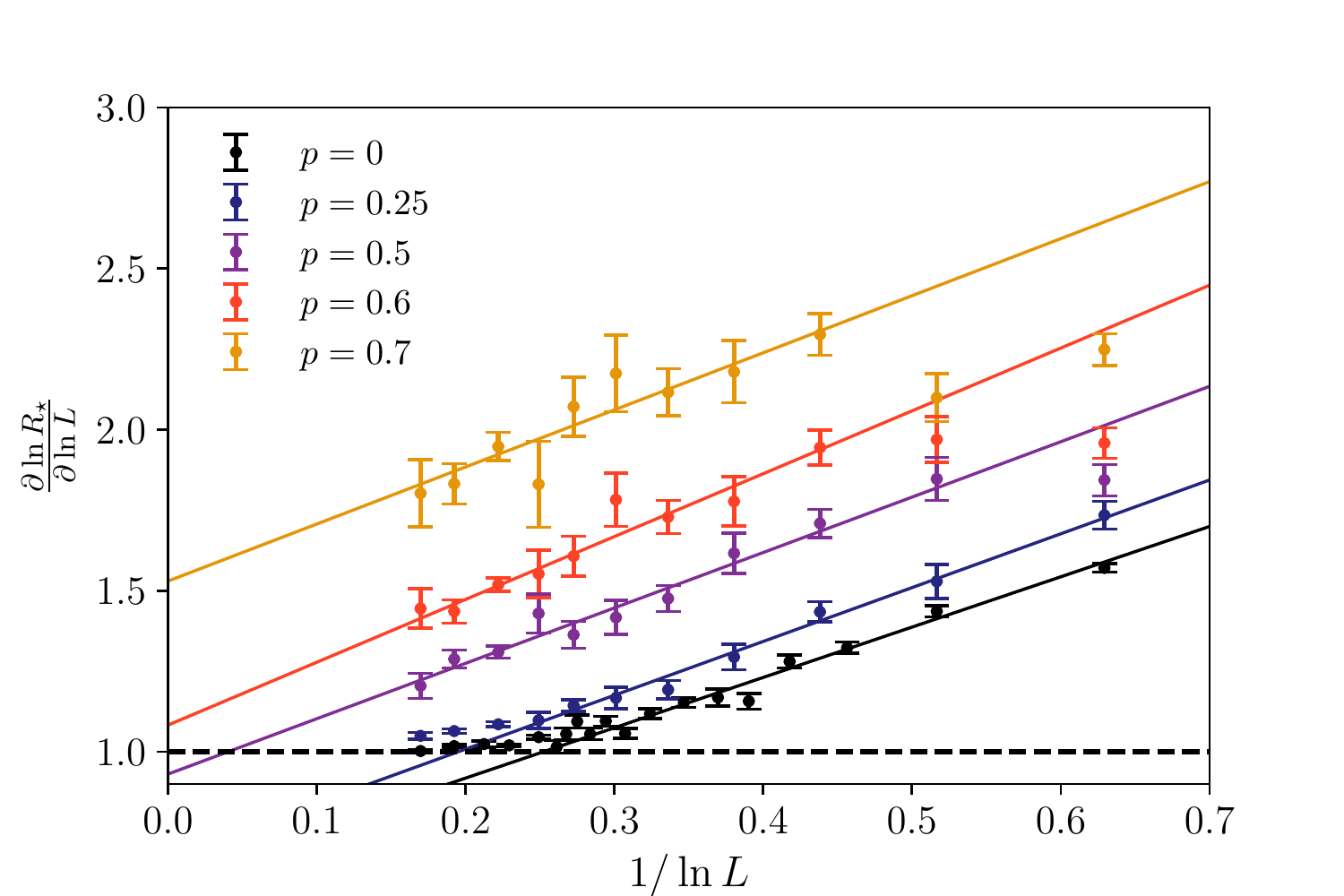}
    \caption{Behavior of the resistance beta function at large $L$, shown for $W = 4$ and several values of $p$ near the diffusion-subdiffusion transition. The lines show fits to the data up to linear order in $1 / \ln L$, with the y-intercept equal to the resistance scaling exponent $\beta$ in the thermodynamic limit.}
    \label{fig:beta_W4}
\end{figure}

To see how $\beta$ increases from 1, it is convenient to look at the discrete beta function, as done previously for the case $p=0$ in Figure \ref{fig:FiniteSize}; here, however, we fix $W$ and change $p$.
In Fig.~\ref{fig:beta_W4} we show the beta function as a function of $1 / \ln L$ for several dephasing fractions $p$ with $W=4$ near the diffusion-subdiffusion transition.
It is clear that the discrete derivative $\beta$ decreases linearly in $1/\ln(L)$ until one of two things happens: it either saturates to $\beta=1$, or it reaches a thermodynamic limit $\beta_\infty>1$. Since the slope of the lines is approximately constant (or at most it changes slowly with $p$: for $W=4$ it is between $1.6$ and $2.0$), we find that
\begin{equation}
    \beta = \begin{cases}
    1 & \text{if } p \leq p_c, \\
    1+c(p-p_c)+O((p-p_c)^2) & \text{if } p > p_c.
    \end{cases}
\end{equation}
For $W=4$ we find $p_c=0.54\pm 0.01$, and the coefficient $c\simeq 1.8$. Similarly, for $W=3$ we find $p_c = 0.667 \pm 0.007$ and for $W=2$ we find $p_c = 0.830 \pm 0.001$.
These results are consistent with those found from the histogram tail exponent $\nu$ (as can be seen from the comparison shown in the lower panel of Fig.~\ref{fig:PLtails}).

The critical exponent $(p-p_c)^1$, which is also observed in the interacting model \cite{Znidaric2016Diffusive}, looks typical of a mean-field scenario. We present and solve a semiclassical model of subdiffusion in Section~\ref{sec:classical}; there we also find a critical exponent of one, and we can determine $c$ explicitly in terms of the microscopic parameters of the model.

The last thing one can extract from this analysis is a critical length scale for $p<p_c$, which is the length $L$ at which the curves start bending upward and transport becomes diffusive. This crossover length can be defined, roughly, as the intersection point between the linear extrapolation (as a function of $1/\ln L$) and the line $\beta=1$. One finds:
\begin{equation}
    L\simeq L_0 \exp\left(\frac{b}{p_c-p}\right),
\end{equation}
where, for $W=4$, $p_c=0.54$ as before, $b \approx 1$ and $L_0=O(1)$.
To give an idea of how quickly this function grows, consider that for $p=0.3$ we have $L \simeq 1,600$, while for $p=0.4$ we find $L\simeq 10^6$. Notice that this is in line with the more complex Griffiths scenarios given by strong disorder renormalization group (SDRG) \cite{Vosk:2013kq,altman2015review,Gopalakrishnan2020Dynamics}, which predict an infinite dynamical exponent $z$ (according to the definition $L\sim (p_c-p)^{-z}$).

\section{Exactly solvable semiclassical model}
\label{sec:classical}
 
In order to understand the results shown in the previous section, we now examine a related phenomenological semiclassical model.
Consider a chain of $L$ units, where each unit may be either an insulator with probability $p$, or a conductor with probability $1-p$.
Conductors combine linearly, with each conductor contributing a resistance of $R_0$, so a string of conductors of length $s$ has a resistance of $R_0 s$. 
Because of phase coherence, insulators combine multiplicatively, so a string of insulators of length $s$ has a resistance of $R_1 q^s$, where $q>1$.
The total resistance of the chain is then equal to
\begin{equation}
\label{eq:R}
R = R_0 \sum_{s \geq 1} n_c(s) s + R_1 \sum_{s \geq 1} n_i(s) q^s,
\end{equation}
where $n_c(s)$ ($n_i(s)$) is the number of strings of conductors (insulators) of length $s$.
The relationship with the microscopic dephasing model is as follows: strings of sites without dephasing are modelled by strings of insulators (a shorter localization length due to stronger disorder corresponds to a larger $q$), and strings of sites with dephasing  are modelled by strings of conductors.
This semiclassical model is equivalent to the system of random resistors introduced in Ref.~\onlinecite{AgarwalAnomalousDiffusion}.

\subsection{Analytical results}

We will now analyse the statistical properties of the total resistance $R$.
In order to determine the average resistance across configurations, we note that the quantities $n_c(s)$ and $n_i(s)$ are Poisson-distributed random variables:
\begin{equation}
    P \left( n(s) \right)=\frac{\mu(s)^{n(s)}}{n(s)!}e^{-\mu(s)},
\end{equation}
where $\mu=\langle n(s) \rangle$, with the angled brackets denoting an average over different configurations of conductors and insulators.
This is subject to the constraint:
\begin{equation}
    \label{eq:constraint}
    \sum_{s\geq 1}s n_c(s)+\sum_{s\geq 1}s n_i(s)=L.
\end{equation}
The mean values are
\begin{equation}
\begin{aligned}
    \mu_c(s) = \langle n_c(s) \rangle & = & L p^2(1-p)^s,\\
    \mu_i(s) = \langle n_i(s) \rangle & = & L (1-p)^2p^s.
\end{aligned}
\end{equation}
The constraint \eqref{eq:constraint} is then satisfied on average for $L \gg 1$:
\begin{equation}
    \sum_{s\geq 1}\langle s n_c(s)+s n_i(s) \rangle = L (1-p) + L p = L.
\end{equation}
The average resistance is therefore given by
\begin{equation}
    \label{eq:Rav}
    \langle  R \rangle = L R_0 (1-p)+L R_1 \sum_{s\geq 1}(1-p)^2 (q p)^s.
\end{equation}
When $pq < 1$ the sum \eqref{eq:Rav} converges, and the average total resistance grows linearly with system size, meaning the system is diffusive:
\begin{equation}
\begin{aligned}
    \langle R \rangle & = R_0 (1-p) L + R_1 \frac{p q (1-p)^2}{1 - pq} L\\
    & = R'_0 L.
\end{aligned}
\end{equation}
On the other hand, for $pq \geq 1$, the sum \eqref{eq:Rav} does not converge and the average does not exist.
In this regime, the total resistance for a given configuration is dominated by the longest string of consecutive insulators, which has a typical length of $s_{\star} \approx \log_{1/p} (L [1-p]) \approx \log_{1/p} (L)$ for large $L$ \cite{Gordon1986Extreme}.
This then results in a typical resistance of
\begin{equation}
    \label{eq:subdiff}
    R \sim R_1 q^{\log_{1/p} (L)} \propto L^{\log_{1/p} (q)},
\end{equation}
with a subdiffusive scaling exponent $\beta = \log_{1/p} (q)$.
The system therefore has a transition from a diffusive phase to a subdiffusive phase at $p_c = 1/q$:
\begin{equation}
    \beta = \begin{cases}
    1 & \text{if } p \leq p_c = 1 / q, \\
    \log_{1/p} (q) & \text{if } p > p_c.
    \end{cases}
\end{equation}
Close to the transition on the subdiffusive side it follows that
\begin{equation}
    \beta \simeq 1 + \frac{q}{\ln q} (p - p_c) + O \left( (p - p_c)^2 \right), 
\end{equation}
indicating a critical exponent of 1.
In the subdiffusive phase we expect the total resistance to be distributed according to \eqref{eq:RDist}, as the arguments leading to this expression are identical to those described above.
Therefore we expect the subdiffusive phase to be described by the physics of Griffiths effects, with heavy-tailed resistance distributions: $P(R) \propto R^{-1-1/\beta}$ (note that in this phenomenological model the resistances of the insulating clusters are always exponentially large in their size, so the finite-size effects leading to \eqref{eq:StretchExp} do not apply).

We now examine the properties of the distributions of $R$ more carefully and show that this is true.
The Laplace transform of the distribution of the insulating part of the resistance $R_i = \sum_{s=1}^L n_i(s) q^s$ (i.e.\ its moment generating function) is equal to:
\begin{equation}
\begin{aligned}
    \left\langle e^{- \rho R_i} \right\rangle & = \prod_{s=1}^L \left\langle e^{-\rho n_i(s) q^s} \right\rangle\\
    & = \prod_{s=1}^{L}\exp \left( -\mu_i(s) + \mu_i(s) e^{- \rho q^s} \right)\\
    & \equiv e^{- \phi(\rho)},
\end{aligned}
\end{equation}
where the second line follows from evaluating the average for the single $n_i(s)$ with $L \to \infty$. 
The cumulant generating function for $R_i$ is therefore given by $\phi (\rho) = \sum_{s=1}^L \mu_i(s) (1 - e^{-\rho q^s})$.
Examining the lowest few cumulants we find:
\begin{equation}
\begin{aligned}
    \phi(0) & = 0\\
    \langle R_i \rangle = \phi'(0) & = \sum_{s=1}^L \mu_i(s) q^s = L (1-p)^2 \sum_{s=1}^L (p q)^s\\
    \langle \delta R_i^2 \rangle = -\phi''(0) & = \sum_{s=1}^L \mu_i(s) q^{2s} = L (1-p)^2 \sum_{s=1}^L (p q^2)^s\\
    \langle \delta R_i^3 \rangle = \phi'''(0) & = \sum_{s=1}^L \mu_i(s) q^{3s} = L (1-p)^2 \sum_{s=1}^L (p q^3)^s
\end{aligned}
\end{equation}
where $\delta R_i = R_i - \langle R_i \rangle$.
If each sum converged as $L \to \infty$, then the distribution would have a limit where every cumulant is proportional to $L$.
However, for $q > 1$ and any $p > 0$ there always exists an $n$ such that $p q^n > 1$.
Defining $\tau = \log_{1/p} (q)$, the smallest integer larger than $1/\tau$ corresponds to the lowest cumulant that scales superlinearly with $L$, and all subsequent moments will scale with a different power of $L$ (note that $\beta = \tau$ in the subdiffusive phase).
In other words, when $p > q^{-n}$ the $n$\textsuperscript{th} cumulant stops growing linearly with $L$, and begins to scale like $L^{n \tau}$.

\begin{figure}
    \centering
    \includegraphics[width=\columnwidth]{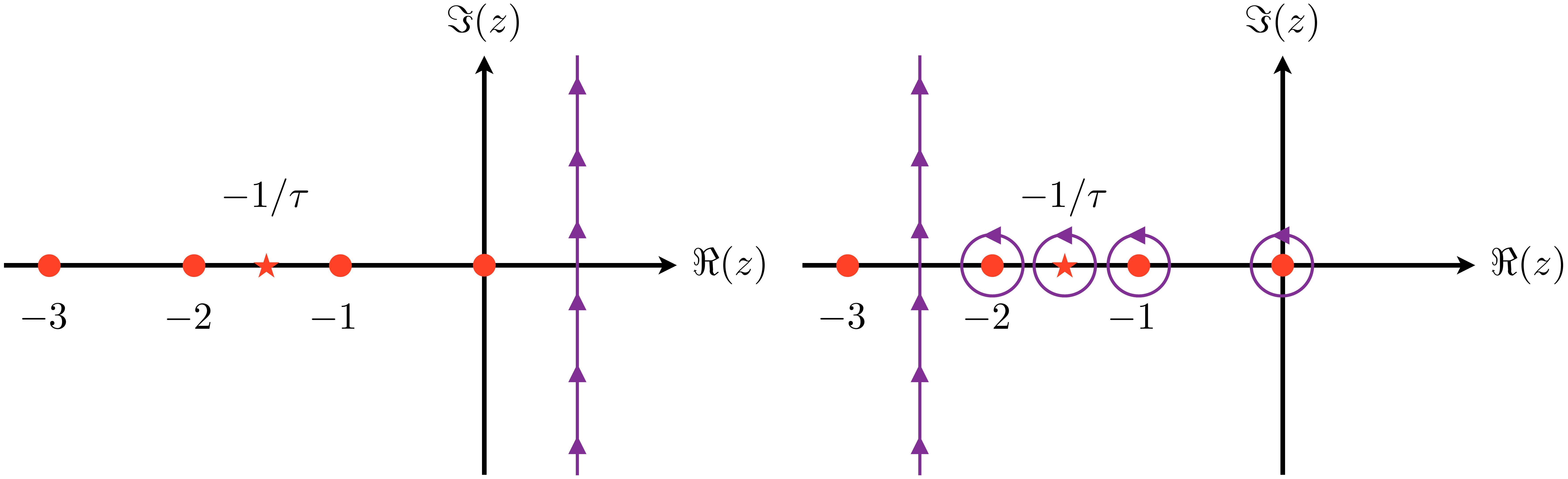}
    \caption{The contours used to evaluate the integral in \eqref{eq:InvMellin} for small $\rho$. Poles of the integrand are indicated in red (poles of the Gamma function as circles and the other pole as a star) and the integration contour is indicated in purple. Left: The Bromwich contour used in \eqref{eq:InvMellin}. Right: Deformation of the contour by pushing it to the left, picking up the leading orders in $\rho$ as each pole is enclosed.}
    \label{fig:contour}
\end{figure}

To analyse the distribution of $R_i$, we extend the sum to $L \to \infty$, therefore neglecting terms exponentially small in $L$:
\begin{equation}
\begin{aligned}
    \phi (\rho) &= \sum_{s=1}^{\infty} \mu_i(s) - L (1-p)^2 \sum_{s=1}^{\infty} p^s e^{- \rho q^s}\\
    &= L (1-p) - L (1-p)^2 \psi (\rho),
\end{aligned}
\end{equation}
where
\begin{equation}
    \psi (\rho) \equiv \sum_{s=1}^{\infty} p^s e^{- \rho q^s}.
\end{equation}
We evaluate the sum after taking the Mellin transform:
\begin{equation}
\begin{aligned}
    M (z) &= \int_0^{\infty} \mathrm{d} \rho \rho^{z-1} \psi(\rho) = \sum_{s=1}^{\infty} p^s \Gamma (z) q^{-sz}\\
    &= \Gamma (z) \frac{1}{p^{-1}q^z - 1},
\end{aligned}
\end{equation}
where $\Gamma(z)$ is the gamma function.
The inverse transform therefore gives
\begin{equation}
    \label{eq:InvMellin}
    \psi (\rho) = \int_{c - i \infty}^{c + i \infty} \frac{\mathrm{d}z}{2 \pi i} \rho^{-z} \Gamma (z) \frac{1}{p^{-1} q^z -1},
\end{equation}
where the integration path is the Bromwich contour $\mathcal{B}$ shown in the left panel of Fig.~\ref{fig:contour}.
The expansion for small $\rho$ can be obtained by moving the contour of the $z$-integration to the left (see the right panel of Fig.~\ref{fig:contour}), picking up the leading-order terms with each pole.
The gamma function has simple poles on all the negative integers, with the pole at $-m$ giving a contribution of $\frac{(- \rho)^m}{m! (p^{-1} q^{-m}-1)}$ to the integral.
There is another simple pole located at $z = -1 / \tau < 0$, which gives a contribution of $\rho^{1 / \tau} \Gamma(-1 / \tau) \ln (q)$ (there is also a sequence of image poles at $z = -1/\tau + 2 \pi i n / \ln (q)$ for $n \in \mathbb{Z}$, however, their contribution is strongly suppressed by their distance from the real line for reasonable values of $q \lesssim 10$).

The leading-order terms depend on the value of $\tau$, resulting in several regimes.
If $1 / \tau > 2$ we find:
\begin{equation}
\begin{aligned}
\phi (\rho) &= \frac{L (1-p)^2 p q}{1 - p q} \rho - \frac{1}{2} \frac{L (1-p)^2 p q^2}{1 - p q^2} \rho^2 + o (\rho^2)\\
&= \langle R_i \rangle \rho - \frac{1}{2} \langle \delta R_i^2 \rangle \rho^2 + o (\rho^2);
\end{aligned}
\end{equation}
stopping at quadratic order, we recognize the cumulant generating function of a Gaussian distribution:
\begin{equation}
    P(R_i) = \int_{\mathcal{B}} \frac{\mathrm{d}\rho}{2 \pi i} e^{\rho R_i - \phi(\rho)} = \frac{1}{\sqrt{2 \pi \langle \delta R_i^2 \rangle}} e^{- \frac{\left(R_i - \langle R_i \rangle \right)^2}{2 \langle \delta R_i^2 \rangle}}.
\end{equation}
However, if $1 < 1/\tau < 2$ the pole at $z = -1/\tau$ contributes, giving:
\begin{equation}
    \phi (\rho) = \langle R_i \rangle \rho -\frac{L (1-p)^2 \Gamma(-\tau^{-1})}{\ln(q)} \rho^{1 / \tau} + o (\rho^{1 / \tau}).
\end{equation}
Stopping at this order, we recognize the result as consistent with a L\'{e}vy alpha-stable distribution with average $\langle R_i \rangle$, and a scale that grows as $\delta R_i^2 \propto L^{2 \tau} \gg L$.
The stability parameter is equal to $1 / \tau$, resulting in a distribution with a tail decaying asymptotically as $P(R_i) \propto R_i^{-1 - 1 / \tau}$.
If $1 / \tau < 1$ then the distribution has a heavy tail and the average no longer exists, so we must instead consider the typical value of $R_i$.
Noting that in this regime $\beta = \tau$, we recognize the heavy-tailed distribution from the Griffiths effects argument, $P(R_i) \propto R_i^{-1 - 1 / \beta}$, resulting in the scaling $R \approx R_i \propto L^{\beta}$.

\subsection{Numerical results}
 
We now study the system numerically in order to confirm the accuracy of the analysis above.
The results shown correspond to the parameters: $R_0=1$, $R_1=1.5$, and $q=1.5$.
By changing $p$ we can tune the system across the diffusion-subdiffusion transition, which should be found at $p_c = 2/3$.
We introduce additional randomness by making $R_i(s)$ the product of $s$ random variables $q_n$, each drawn independently from a narrow uniform distribution in the range $q \pm 0.1$, so $R_i(s) = R_1 \prod_{n=1}^s q_n$ (note that on average $R_i(s)$ is still equal to $R_1 q^s$); this has no effect on the results other than to smooth the histograms of $R$.
The numerical analysis of this model requires no solution of matrix equations, unlike the microscopic model in Section~\ref{sec:dephasing}, allowing us to extensively sample very large system sizes: the results below include systems of up to $L \sim 10^6$ and $\sim 10^5$ samples.

\begin{figure}
    \centering
    \includegraphics[width=\columnwidth]{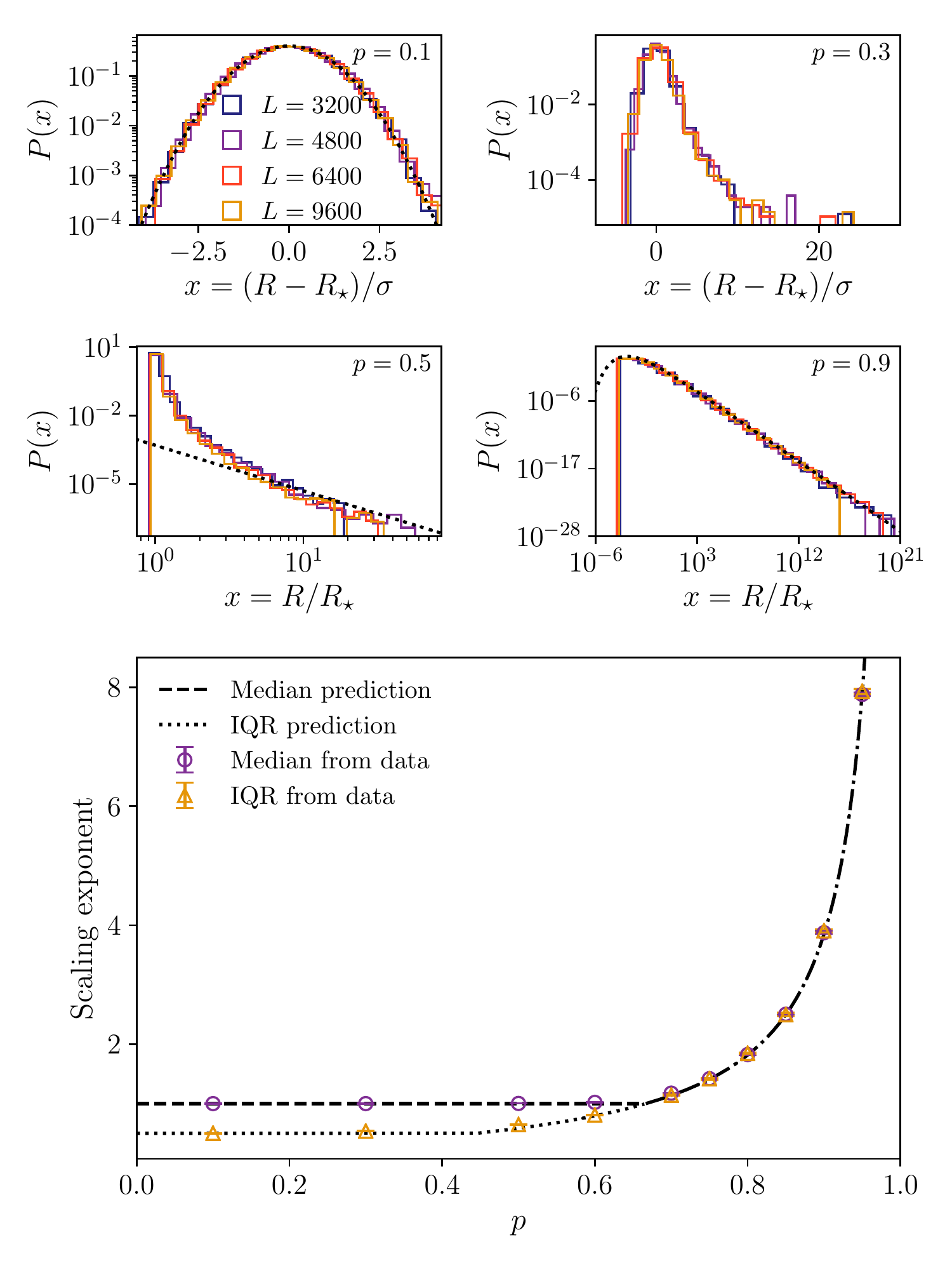}
    \caption{Comparison of analytical and numerical results for the semiclassical model.
    Upper panels: Histograms of the rescaled resistance in the diffusive phase $p=0.1$ (top left), $p=0.3$ (top right), and $p=0.5$ (center left) and the subdiffusive phase $p=0.9$ (center right). The top panels show the rescaling $(R - R_{\star}) / \sigma$, which is compared to a Gaussian distribution (black dotted line) for $p=0.1$, the existence of a tail in the diffusive phase is clear in the $p=0.3$ histogram.
    The central panels show the rescaling $R / R_{\star}$, with a comparison to the predicted tail for $p=0.5$, and a comparison to the theoretical prediction \eqref{eq:RDist} shown for $p=0.9$ (black dotted lines).
    Lower panel: Comparison of the numerically determined scaling exponent for the median resistance, $\beta$, and also for the inter-quartile range, $\beta'$, with the analytical prediction $\log_{1/p} (q)$. The critical point is $p_c=1/q=2/3$.
    }
    \label{fig:classical}
\end{figure}

A comparison between the numerical results and the analytical predictions is shown in Fig.~\ref{fig:classical}, using results on systems of up to $L = 128,000$.
The lower panel shows a good agreement between the median resistance scaling exponent $\beta$ and the theoretical prediction: $\beta = 1$ for $p < p_c = 2/3$ and $\beta = \log_{1/p} (q)$ for $p > 2/3$.
We also see that the width of the distributions, measured by the inter-quartile range (IQR), also scales as expected: for $\text{IQR} \propto L^{\beta'}$ we find $\beta' = 1/2$ for $p < q^{-2} = 4/9$ and $\beta' = \log_{1/p} (q)$ for $p > 4/9$.
The numerical values of $\beta$ and $\beta'$ were extracted from fits to the data of the form $a L^{\beta}$ using $L \geq 1600$ (near $p_c$, where finite-size effects are strongest, we used $L \geq 6400$).
The predicted values for $\beta$ and $\beta'$ are indicted by the dashed and dotted black lines respectively (the line becomes dot-dashed for $p > p_c$, where $\beta = \beta'$). 
The upper panels show histograms of the resistance over a range of $p$ values.
Deep in the diffusive phase ($p = 0.1$ in the figure) the distribution is approximately Gaussian (indicated by the black dotted line), with the average and median growing linearly with $L$ and the standard deviation growing like $\sqrt{L}$.
Closer to the subdiffusion transition ($p = 0.3$ in the figure) the distribution starts to develop a tail (these parameters correspond to the point where the third cumulant has started to scale faster than linearly with $L$, $p q^3 = (0.3)(1.5)^3 = 81/80$).
Close to the subdiffusion transition on the diffusive side ($p=0.5$ in the figure, where the average is still defined but the variance is not, $p q^2 = (0.5)(1.5)^2=9/8$) we see that the distribution has developed the predicted weak power-law tail which is indicated by the black dotted line.
In the subdiffusive phase ($p = 0.9$ in the figure) the distribution has a strong power-law tail, which agrees well with the prediction \eqref{eq:RDist} that $R$ is dominated by the longest string of insulators, shown again by the black dotted line.
The discrepancy at small $R$ is due to the fact that these realizations have unusually short longest strings of insulators, which therefore have a less dominant contribution to the total resistance.

\begin{figure}
    \centering
    \includegraphics[width=\columnwidth]{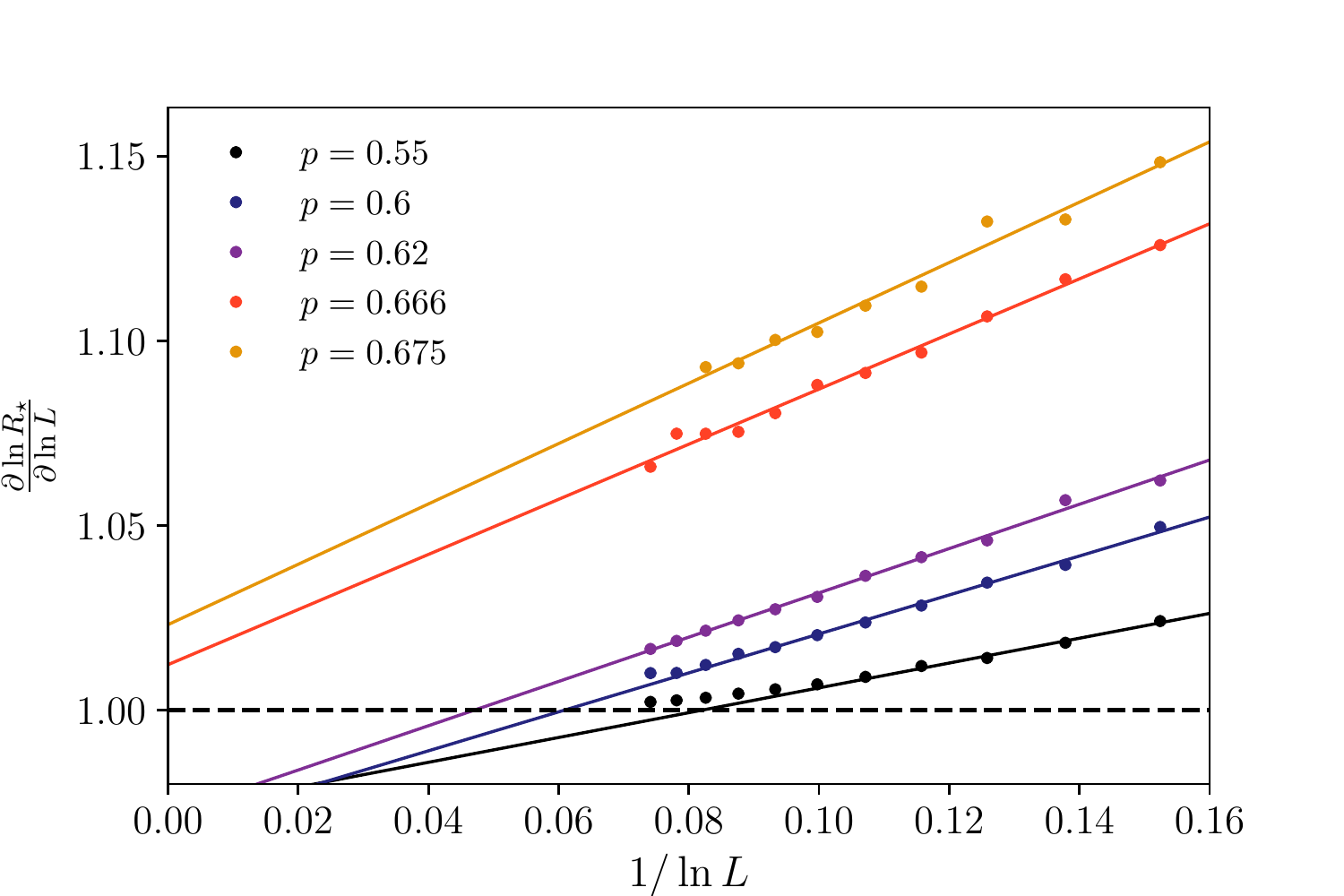}
    \caption{The resistance beta function for the semiclassical model, plotted against $1 / \ln L$ for a range of $p$ values. The lines indicate fits to linear order in $1 / \ln L$.}
    \label{fig:BetaClassical}
\end{figure}

In Fig.~\ref{fig:BetaClassical} we show the discrete resistance beta function for the semiclassical model, plotted as a function of $1 / \ln L$ for comparison with the results from the dephasing model shown in Fig.~\ref{fig:beta_W4}.
In this plot we have collected data for very large systems, up to $L = 1,024,000$, in order to examine the finite-size effects, and we see that the discrete derivative appears to decrease linearly with $1 / \ln L$, as was the case for the dephasing model. The critical point is $p_c=1/q=2/3=0.666$. From the figure one can see that a linear fit in $1/\ln L$ for $L$ up to $10^6$ still gives an error of about $1\%$ in the asymptotic value of $\beta$ (i.e.\ $\beta=1.01$ instead $1.00$) and this implicates a comparable error in the critical value $p_c$. On the other hand, a $1/\ln L$ dependence means that nothing much changes if $L$ is considerably \emph{reduced}, and so a few $\%$ error on the asymptotic quantities comes from considering $L=O(10^3)$ (which are the kind of system sizes amenable to TEBD numerics \cite{Znidaric2016Diffusive,Schulz2020Phenomenology}).

\section{Discussion}
\label{sec:Disc}

In this paper we have studied DC spin transport in a disordered, non-interacting spin chain with dephasing on random sites.
Using this model we can study transport in a system with insulating and thermal regions at much larger system sizes than is possible for interacting models (even when employing powerful matrix-product operator methods).
We have shown that the system exhibits a phase transition from diffusive to subdiffusive transport when the density of sites with dephasing decreases below a critical value.
In the subdiffusive phase the distributions of resistances across different realizations of the disorder and dephasing have heavy tails, suggesting that the subdiffusion is caused by Griffiths effects.

We have also presented a related, exactly solvable semiclassical model, where the system is formed of randomly chosen sequences of insulators and conductors.
We have shown that this system also undergoes a transition from diffusion to subdiffusion due to Griffiths effects when the density of conductors decreases below a critical value.
This model captures the qualitative features seen in the microscopic quantum model, including the Gaussian distributions of resistances deep in the diffusive phase, which develop tails as the transition to subdiffusion is approached, and eventually become heavy-tailed in the subdiffusive phase.

The behavior of the quantum model is most similar to that of the semiclassical model (i.e.\ most consistent with the physics of Griffiths effects) when the disorder is strong and the subdiffusion weak.
We have argued that this discrepancy is due to the finite lengths of the clusters of sites with and without dephasing: the semiclassical model is constructed using the asymptotic scaling properties of these clusters, and we have shown that for certain parameter combinations they are certainly not in their asymptotic regimes.
At very large system sizes we expect that the behavior of the two models will become increasingly similar.

\begin{figure}[t]
    \centering
    \includegraphics[width=\columnwidth]{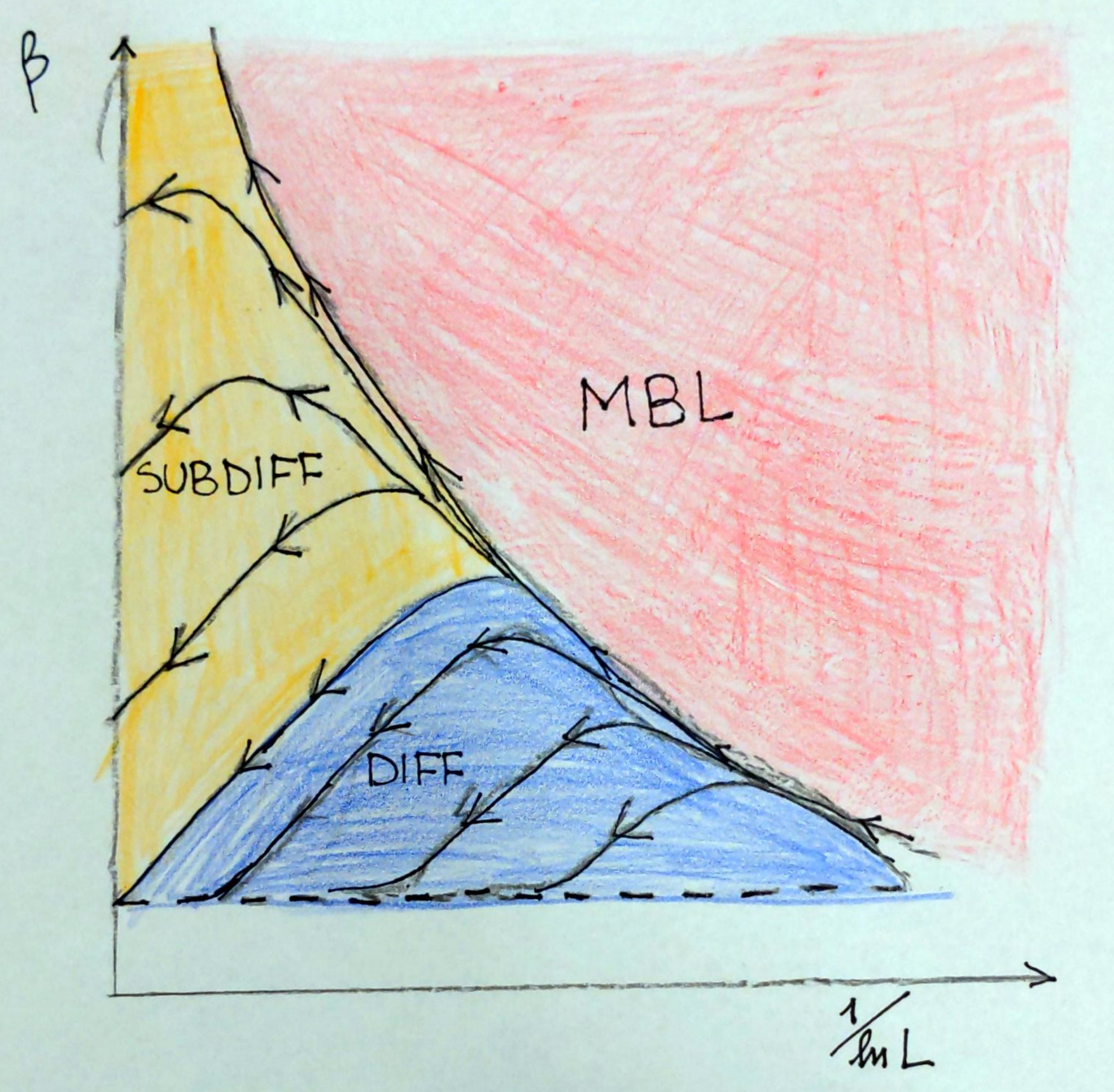}
    \caption{The dynamical phase diagram of the random dephasing model and the semiclassical model, showing a schematic of the finite-size flow of the beta function. For $p < p_c$ the system is diffusive, with $\beta \to 1$ in the termodynamic limit, while for $p > p_c$ the system is subdiffusive with $\beta > 1$. In the limit $p \to 1$ the system is localized and $\beta \propto L$ diverges in the thermodynamic limit. Above this line, if it existed for this model, would lie the MBL phase, where one has the very fast divergence $\beta=\frac{1}{\xi}\exp(1/[1/\ln L])$ with $\xi$ the many-body localization length.}
    \label{fig:phasediagram}
\end{figure}

After looking at sufficiently many figures for the beta function $\partial\ln R/\partial\ln L$, one observes that the flows for different $p$ values do not interesect. This is a signature that $\ln R$ is indeed a good scaling function, and one can discuss an underlying renormalization group of sorts (most probably a kind of SDRG). Therefore, one can infer a flow diagram for such an RG. In Fig.~\ref{fig:phasediagram} we show such a phase diagram for the random dephasing model and the semiclassical model, based on a schematic of the finite-size flow of the beta function, summarizing the results of Sections~\ref{sec:dephasing} and \ref{sec:classical}.
As described above, in the diffusive phase the beta function always flows to $\beta = 1$, while in the subdiffusive phase $\beta > 1$ in the thermodynamic limit.
For large $p$ and small enough $L$ the system is effectively an insulator, with a resistance scaling like $R \propto q^L$, resulting in a beta function that increases linearly with $L$.
When $L$ increases above a length scale that grows like $(1-p)^{-1}$ this growth reverses and the beta function begins to decrease towards its asymptotic value (approximately linearly in $1 / \ln L$, as described above).
Only for $p = 1$ will the growth of the beta function continue all the way to the thermodynamic limit, signalling localization. If an MBL region existed in this model, it would lie in the pink region, as indicated.

The physics of these models is relevant to our understanding of subdiffusion in interacting quantum systems, which is also believed to be caused by the presence of rare insulating regions.
Presumably in interacting systems, if the regions of strong disorder are not large enough to act as bottlenecks to transport, the nature of the subdiffusion may be concealed by finite-size effects similar to those described in Section~\ref{sec:finitesize}: the subdiffusive scaling of $R_{\star}$ with $L$ in the absence of heavy-tailed distributions, as seen for $W=1$ in Fig.~\ref{fig:PLtails}, is reminiscent of the results in Ref.~\onlinecite{Schulz2020Phenomenology}.
In this paper we have demonstrated in a fully quantum mechanical model how these heavy-tailed distributions can be hidden by the unconventionally slow finite-size flow.
This offers a potential reconciliation of the results in Ref.~\onlinecite{Schulz2020Phenomenology} with that of \cite{Agarwal2017Rare} (i.e.\ the predicted distributions may have been observed if it were possible to study the interacting systems at sizes such as those investigated here for the random dephasing model).

Determining the sizes of the rare insulating regions in interacting models, and how this affects their effectiveness as bottlenecks, would be an important step in confirming or refuting the Griffiths effects hypothesis, and this could potentially be achieved using probes of local thermal properties such as those employed in Refs.~\onlinecite{Lenarcic2018Activating,Lenarcic2019Critical}. It could also be enlightening for the theory of the transition, helping in supporting and discriminating between the various renormalization groups scenarios \cite{khemani2017critical,dumitrescu2019kosterlitz, goremykina2019analytically,morningstar2020many,thiery2017microscopically,laflorencie2020chain} which have been proposed and which lead to different critical properties of the dynamical MBL transition. 
 
\acknowledgements
The authors would like to thank Sarang Gopalakrishnan, Vadim Oganesyan, Vipin Kerala Varma, and Marko \v{Z}nidari\v{c} for insightful discussions and collaborations in the early stages of this project. We would also like to thank Carlotta Scardicchio for making Fig.~\ref{fig:phasediagram}. This work was supported by the Trieste Institute for the Theory of Quantum Technologies.
 
\bibliographystyle{apsrev4-1}
\bibliography{MBLbib}
 
\appendix

\section{Finding the NESS current}
\label{app:method}

The NESS current corresponding to the system described by equations \eqref{eq:H}-\eqref{eq:dephasing} can be calculated exactly, and here we briefly outline the method to do so.
Detailed discussions and derivations of these equations can be found in Refs.~\onlinecite{Prosen2008Third,Znidaric2010Exact,Znidaric2013Transport}.
We use the correlation matrix in the NESS to calculate the quantities of interest, namely the expectation values of the onsite magnetization $\sigma^z_n$ and the current through the bond leaving site $n$ in the positive direction $j_n = 2 \left( \sigma^x_n \sigma^y_{n+1} - \sigma^y_n \sigma^x_{n+1} \right)$ (which can be derived from the continuity equation for the local magnetization).
The correlation matrix is an $L \times L$ matrix from which we can calculate our quantities of interest: $\langle \sigma^z_n \rangle = -C_{n,n}$ and $\langle j_n \rangle = 4 \, \Im \left( C_{n,n+1} \right)$, where the NESS current $j_{\infty}$ should be independent of $n$.

The correlation matrix is found by numerically solving the matrix equation
\begin{equation}
    \label{eq:MatEq}
    A C + C A^{\dagger} + G \tilde{C} + \tilde{C} G = P,
\end{equation}
where $\tilde{C}$ is the correlation matrix with the diagonal elements removed (note that for uniform dephasing, $\gamma_n = \gamma$, this reduces to equation 11 of Ref.~\onlinecite{Znidaric2013Transport}).
The non-Hermitian matrix $A = i E - i J + \Gamma R$, where $E_{n,n} = h_n$, $J_{n,n \pm 1} = -1$ (note that $E-J$ is the Hamiltonian \eqref{eq:H} in the single-particle sector), and $R_{1,1} = R_{L,L} = 1$.
The remaining matrices are: $-P_{1,1} = P_{L,L} = 2 \Gamma$ and $G_{n,n} = \gamma_n$.
All unspecified matrix elements are zero.\\

When the disorder is strong and the system is large, the current $j_{\infty}$ becomes small and imperfect numerical precision can result in the solution of \eqref{eq:MatEq} being unphysical.
This is easily diagnosed by studying the properties of the solution, such as the spatial invariance of the current, and whether the magnetization profile is real and bounded by $-1 \leq \langle \sigma^z_n \rangle \leq 1$.
An alternative method of solving \eqref{eq:MatEq} was presented in Ref.~\onlinecite{Varma2017Fractality} for a system without dephasing, which can be generalized to a system where $\gamma \neq 0$.
Defining the non-Hermitian matrix $T = A + G$, we numerically find its complex eigenvalues $\lambda_n$ and left and right eigenvectors $| \psi^{(Ri)} \rangle = \sum_k \psi^{(Ri)}_k | k \rangle $ and $| \psi^{(Li)} \rangle = \sum_k \psi^{(Li)}_k | k \rangle$, normalized such that $\langle \psi^{(Lm)} | \psi^{Rn} \rangle = \delta_{m,n}$.
The eigenvectors are complex conjugates of each other $\psi^{(Li)}_k = \left( \psi^{(Ri)}_k \right)^*$.
We can rewrite \eqref{eq:MatEq} as:
\begin{equation}
    \label{eq:MatEq2}
    T C + C T^{\dagger} = \tilde{P},
\end{equation}
where $\tilde{P}$ is a diagonal matrix with elements equal to $\tilde{P}_{n,n} = P_{n,n} + 2 \gamma_n C_{n,n}$, which has the formal solution:
\begin{equation}
   C = \int_0^{\infty} \mathrm{d}t \; e^{- t T} \tilde{P} e^{- t T^{\dagger}}.
\end{equation}
Rewriting this in the eigvenbasis of $T$ and evaluating the integral we find a set of equations:
\begin{equation}
    \label{eq:SpectralC}
    C_{m,n} = \Gamma \left( \theta_{m,n,1} - \theta_{m,n,L} \right) + \sum_k \gamma_k \, \theta_{m,n,k} \, C_{k,k},
\end{equation}
where we have introduced the shorthand:
\begin{equation}
    \theta_{j,k,l} = 2 \sum_{p,q} \frac{\psi^{(Rp)}_j \psi^{(Rp)}_l \left( \psi^{(Rq)}_k \psi^{(Rq)}_l \right)^*}{ \lambda_p + \lambda_q^* }.
\end{equation}
After numerically diagonalizing the matrix $T$ the quantities $\theta_{j,k,l}$ can be constructed easily.
The diagonal elements, which describe the magnetization profile $\langle \sigma^z_n \rangle = -C_{n,n}$, give us a set of $L$ linear equations:
\begin{equation}
     \langle \sigma^z_n \rangle = \Gamma \left( \theta_{n,n,1} - \theta_{n,n,L} \right) + \sum_m \gamma_m \, \theta_{n,n,m} \, \langle \sigma^z_m \rangle,
\end{equation}
which can be solved numerically.
With the knowledge of the diagonal elements we can then simply evaluate the currents:
\begin{equation}
\begin{aligned}
    \langle j_n \rangle = 4 \Im \Bigg( \Gamma & \left[ \theta_{n,n+1,L} - \theta_{n,n+1,1} \right]\\
    & - \sum_m \gamma_m \, \theta_{n,n+1,m} \,\langle \sigma^z_m \rangle \Bigg).
\end{aligned}
\end{equation}
Note that in order to calculate the magnetization profile and the current through every bond it is not necessary to evaluate all $L^3$ of the $\theta_{j,k,l}$ quantities.

\end{document}